\documentclass[journal]{IEEEtran}
\usepackage{amssymb}
\usepackage{mathrsfs}
\usepackage{graphicx}
\usepackage{amsmath, amssymb, amsthm}
\usepackage{leftidx}
\usepackage{extarrows}
\usepackage{stfloats}
\usepackage{picinpar}
\usepackage{enumerate}
\usepackage{algorithm}
\usepackage{algorithmicx}
\usepackage{algpseudocode}
\usepackage{epsfig}
\usepackage{latexsym}
\usepackage{amsfonts}
\usepackage{enumerate}
\usepackage{graphics}
\usepackage{graphicx,subfigure}
\usepackage{MnSymbol}
\usepackage{float}
\usepackage{pict2e}
\usepackage{tikz}
\usepackage{bm}
\newtheorem{theorem}{Theorem}
\newtheorem{assumption}{Assumption}
\newtheorem{corollary}{Corollary}
\newtheorem{lemma}{Lemma}
\newtheorem{remark}{Remark}
\newtheorem{definition}{Definition}
\newtheorem{example}{Example}

\newtheorem{problem}{Problem}

\title{Polynomial-Time Algorithms for Structurally Observable Graphs by Controlling Minimal Vertices}

\author{Shiyong~Zhu, Jianquan Lu$^\ast$,~\IEEEmembership{Senior Member,~IEEE}, Daniel W.C. Ho,~\IEEEmembership{Fellow,~IEEE}, and Jinde Cao,~\IEEEmembership{Fellow,~IEEE}
\thanks{This work was supported by the National Natural Science Foundation of China under Grant No. 61973078 and ''333 Engineering'' Foundation of Jiangsu Province of China under Grant BRA2019260.}\\
\thanks{Corresponding author: Jianquan Lu.}
\thanks{Shiyong Zhu is with the Department of Systems Science, the School of Mathematics, Southeast University, Nanjing 210096, China, and is also with the College of Mathematics and Computer Science, Zhejiang Normal University, Jinhua 321004, China (email: zhusy0904@gmail.com).}
\thanks{Jianquan Lu is with the Department of Systems Science, the School of Mathematics, Southeast University, Nanjing 210096, China (email: jqluma@seu.edu.cn).}
\thanks{Daniel W.C. Ho is with the Department of Mathematics, City University of Hong Kong, Kowloon, Hong Kong (email: madaniel@cityu.edu.hk).}
\thanks{Jinde Cao is with the Department of Systems Science, the School of Mathematics, Southeast University, Nanjing 210096, China, and is also with the Yonsei Frontier Lab, Yonsei University, Seoul 03722, South Korea. (e-mail: jdcao@seu.edu.cn)}
}

\begin{document}
\maketitle
\thispagestyle{empty}
\pagestyle{empty}
\begin{abstract}
  The aim of this paper is to characterize an important class of marked digraphs, called structurally observable graphs (SOGs), and to solve two minimum realization problems. To begin with, by exploring structural observability of large-scale Boolean networks (LSBNs), an underlying type of SOGs is provided based on a recent observability criterion of conjunctive BNs. Besides, SOGs are also proved to have important applicability to structural observability of general discrete-time systems. Further, two minimum realization strategies are considered to induce an SOG from an arbitrarily given digraph by marking and controlling the minimal vertices, respectively. It indicates that one can induce an observable system by means of adding the minimal sensors or modifying the adjacency relation of minimal vertices. Finally, the structural observability of finite-field networks, and the minimum pinned node theorem for Boolean networks are displayed as application and simulation. The most salient superiority is that the designed algorithms are polynomial time and avoid exhaustive brute-force searches. It means that our results can be applied to deal with the observability of large-scale systems (particularly, LSBNs), whose observability analysis and the minimum controlled node theorem are known as intractable problems.
\end{abstract}


\section{Introduction}\label{section-introduction}
Consider the distributed discrete-time system as
\begin{equation}\label{equ-DTS}
\left\{\begin{aligned}
&x_i[k+1]=f_i((x_\alpha[k])_{\alpha\in \mathbf{N}_i}),~i=1,2,\cdots,n\\
&y_j[k]=x_j[k], ~j=1,2,\cdots,h
\end{aligned}\right.
\end{equation}
where each state variable or output variable taking values in domains $\mathbb{R}^n$ or $\mathbb{R}^h$ are described as $x_i[k]$ or $y_j[k]$, and holistic state variable of $n$ agents and the corresponding output measurement are compactly performed as column vectors ${\bf x}[k]=(x_1[k],x_2[k],\cdots,x_n[k])^\top$ and ${\bf y}[k]=(y_1[k],y_2[k],\cdots,x_h[k])^\top$ respectively, where index $k$ captures discrete time instant $k$. Output $y_j[k]$ is taken as a {\em sensor} imposed on agent $j$ because it accurately records the real-time value of agent $j$. Without loss of generality, one can suppose that system sensors always reflect the values of the first $h$ state variables as in (\ref{equ-DTS}). Then, the data transmission among agents, in iteration (\ref{equ-DTS}), can be characterized by a digraph $\mathrm{G}:=(\mathrm{V},\mathrm{E})$ with $\mathrm{V}:=\{v_1,v_2,\cdots,v_n\}$, where vertex $v_i \in \mathrm{V}$ stands for agent $i$, and $(v_i,v_j)\in\mathrm{E}$ if $j \in \mathbf{N}_i$. In this paper, this $n \times n$ digraph $\mathrm{G}$ is generally termed as the {\em interaction digraph} of iteration (\ref{equ-DTS}). In other scenes, like logical network (\ref{equ-DTS}), it is also called {\em dependency graph}.

On account of the diversity of dynamical evolutions, except for multi-agent systems \cite{youky2011tac2262,lin2009auto2154}, system (\ref{equ-DTS}) actually involves many classical networks, including but not limited to, Boolean networks \cite{kauffman1969jtb437,chengdz2011springer,Margaliot2019TAC2727,valcher2015aut21,fagiolini2013auto2339}, finite-field networks \cite{sundaram2012TAC60,pasqualetti2014auto349,mengmin2020auto108877,lixx2016auto39}, as well as shift registers \cite{zhongjh2016ieeetc2274}. To just list an example in detail \cite{fagiolini2013auto2339}, if each agent transmits values from a finite set $\{0,1,2,\cdots,\varrho-1\}$ via pre-assigned logical functions, iteration (\ref{equ-DTS}) is termed as a $\varrho$-valued logical network \cite{guoyq2020tac}. Again, it becomes the well-known Boolean network \cite{kauffman1969jtb437} if $\varrho=2$. For iteration (\ref{equ-DTS}), observability is an elementary but significant problem aiming to uniquely determine the initial condition ${\bf x}[0]$ of system from the observation of output evolution over some time windows. The observability of system (\ref{equ-DTS}) can be defined as follows:
\begin{definition}\label{definition-DTMAS-observability}
System (\ref{equ-DTS}) is said to be observable (within finite time) if the output evolutions of network (\ref{equ-DTS}) starting from arbitrary two different initial states at $k=0$ do not coincide at certain time instant.
\end{definition}

\subsection{Background and Motivations}
No matter which type of system (\ref{equ-DTS}), such as logical networks or finite-field networks, many significant results have been established for system observability on the basis of known node dynamics (see, e.g., \cite{Margaliot2013Observability2351,Valcher2012TAC1390,Margaliot2019TAC2727,zhusy2021tac,sundaram2012TAC60,zhangkz2020tac}). However, these theoretical results are always restricted by certain stringent  conditions. In terms of conventional Boolean networks, all results on the observability (see, e.g., \cite{chengdz2009tac1659,guoyq2019auto230,zhangkz2015IEEETAC2733,Valcher2012TAC1390,Margaliot2013Observability2351,zhusy2021tac}) were based on algebraic state space representation approach proposed by Cheng {\em et al}. \cite{chengdz2011springer}. Nevertheless, the established criteria take at least $O(n2^{n})$ time complexity and need to address a $2^n \times 2^n$-dimensional network transition matrix except those in \cite{Margaliot2019TAC2727}. It results in that these conditions are only suitable for the observability analysis of Boolean networks with $n\leq30$ nodes \cite{zhangkz2020tac}. With regard to controllability, the node number of considered Boolean networks is limited to $25$ if we adopt the algebraic state space representation approach. It is far from practice for large-scale Boolean networks (LSBNs), such as the T-cell receptor signaling model with $90$ nodes, which is known as the largest Boolean network of a cellular network to date \cite{largestBN}. Even if Weiss and Margaliot have established a polynomial-time observable condition in \cite{Margaliot2019TAC2727}, the result is only applicable to conjunctive Boolean networks, but not to those conventional ones coupled by all kinds of logical operators. More importantly, \cite{Azuma2015TCNS179,Azuma2019TCNS464,azuma2019tcbb} investigated the problems via the interaction digraphs rather than pursuing the node dynamics approach. The main reason is that biologists consider that interaction digraphs are less complicated to be determined while the identification of intact node dynamics is challenging and difficult to be obtained. Motivated by aforementioned limitations, it is of practical and theoretical significance to investigate structural observability of Boolean networks with {\em available} interaction digraphs but {\em unknown} node dynamics. As for the observability of finite-field networks, \cite{sundaram2012TAC60} devoted to designing feasible interaction digraphs and assigning network weights to generate an observable network. However, this condition cannot guarantee the observability of finite-field networks by utilizing their interaction digraphs, i.e., strongly structural observability.

Recently, many results on analysis and control for structural networks have appeared via directly considering interaction digraphs (see, e.g., \cite{linct1974tac-structural,Azuma2019TCNS464,jiajj2021tac391,zhusy2020framework,Azuma2015TCNS179}). Analyzing system behaviors via interaction digraphs is practically useful and theoretically interesting. In addition, the interaction digraphs not only utilizes the easily-identified interaction digraphs instead of minute node dynamics, but are also robust with the inaccuracy of systems dynamics brought from noisy data. The earliest attempt for structural linear time-invariant systems was made by \cite{linct1974tac-structural} about structural controllability. Followed by it, Liu {\em et al.} found that the minimum number of inputs, in order to maintain the full control of a structural linear time-invariant system, is equal to the unmatched vertices in the maximum matching of interaction digraphs \cite{liuyy2011nature167}. Again, similar work in \cite{liuyy2013PNAS} was also established to identify the optimal sensors for structural observability of linear time-invariant systems. In the area of Boolean networks, \cite{margaliot2018auto56} and \cite{Margaliot2019TAC2727} confirmed that the minimal controllability problem and minimal observability problem of conjunctive Boolean networks were respectively proved to be {\bf NP}-hard and solved by a polynomial-time algorithm. As for conventional Boolean networks, although the minimum pinned node problem has been explored by several approaches (see, e.g., \cite{lujq2021cyber373,liuy2020cyber,wanglq2019cyber4444}), they essentially depended on the brute-force search with different degree of improvement. Thus, such exhaustive search requires exponentially-increasing time complexity. Since solving the minimal structural observability problem can reduce the control cost at the lowest level, we are particularly interested in identifying the minimal nodes that need to be observed or controlled in order to fully assure the structural observability of iteration (\ref{equ-DTS}).

\subsection{Contributions of This Paper}
As the full information on node dynamics is usually unavailable or associated with noises, structural observability of iteration (\ref{equ-DTS}) is completely revealed by its interaction digraph, where directly observable vertices are marked by a special color. It means that the structural properties can be completely revealed on these marked digraphs. A type of marked digraphs is called structural observable graph (SOGs) if interaction digraph of iteration (\ref{equ-DTS}) being this type ensures system observability. To this end, structural observability of Boolean networks is proposed as a new concept, and the corresponding necessary and sufficient condition is given in the same form as that in \cite{Margaliot2019TAC2727}. Motivated by \cite{Margaliot2019TAC2727}, we show that the digraphs satisfying Properties $P_1$ and $P_2$, as will be given in Definitions \ref{definition-P1} and \ref{definition-P2}, are further proved to be SOGs. Further, while a given digraph is not an SOG, two classes of minimum realization problems are proposed to induce an SOG from it. One is to mark the minimum number of vertices (i.e., to add sensors on the minimum number of nodes of (\ref{equ-DTS})) to induce an SOG, and the other aims to modify the in-neighbor adjacency relations of minimal vertices (i.e, agents of (\ref{equ-DTS})) to obtain an SOG. Consequently, two polynomial-time algorithms are respectively developed to achieve these two goals. The concept of SOGs also has the implications for the strongly structural observability of finite-field networks as well as designing observers for LSBNs. The main contributions of this paper are fivefold:
\begin{itemize}
  \item[(1)] Compared with the existing results on the observability that depend on the intact node dynamics of discrete-time systems (\ref{equ-DTS}) (see, e.g., \cite{guoyq2019auto230,zhangkz2015IEEETAC2733,zhusy2021tac,Valcher2012TAC1390,Margaliot2013Observability2351}), structural observability of iteration (\ref{equ-DTS}) is analyzed by {\em available} interaction digraphs with {\em unknown} node dynamics. As stated in \cite{Azuma2015TCNS179,Azuma2019TCNS464,azuma2019tcbb}, the interaction digraphs are easier to be identified than the node dynamics, thus problems considered here are closer to practical situations. Additionally, once the interaction digraphs have been given, arbitrary modifications on the evolution dynamics would not affect system observability. Hence, our approach is robust with respect to inaccuracies on dynamical evolutions brought from noisy data;
  \item[(2)] The criterion for SOGs is checkable in time $O(n^2)$ with regard to node number $n$, and two polynomial-time algorithms are developed to search and control the minimum number of vertices in an arbitrarily given digraph for the sake of inducing an SOG. By comparison, the minimal vertex control problem for structurally controllable graphs has been proved to be {\bf NP}-hard \cite{zhusy2020framework}. The developed algorithms avoid the brute-force searches, which are widely utilized in \cite{lujq2021cyber373,liuy2020cyber,wanglq2019cyber4444}. Hence, algorithms designed here are applicable to LSBNs and would not be limited by $30$ nodes as the exhaustive state space approach \cite{chengdz2011springer};
  \item[(3)] The SOGs also have applicability in strongly structural observability of networks over finite fields. Additionally, an observable network over finite field considered in \cite{sundaram2012TAC60} can be further designed. The results in \cite{sundaram2012TAC60} require that the size of considered field should be larger than the largest height of spanning forests. While using SOGs, the requirement in \cite{sundaram2012TAC60} can be relaxed if the considered interaction digraphs contains an SOG;
  \item[(4)] Our previous paper \cite{zhusy2020novel} developed a polynomial-time algorithm to search the pinned nodes of Boolean networks, but the number of chosen nodes has not been minimized. By means of solutions to the second minimum realization problem, the pinned nodes can be reselected at the least level from viewpoint of interaction digraphs, in order to consider the least nodes.
  \item[(5)] This paper illustrates that digraphs proposed in\cite{Margaliot2019TAC2727} can have more general applicability to the observability of a very general system (1). Compared with \cite{Margaliot2019TAC2727}, structural observability is proposed as a new concept for Boolean networks (with all types of logical operators) instead of conjunctive ones. Besides, an alternative procedure is developed to induce an SOG by adjusting the network structures, while \cite{Margaliot2019TAC2727} focuses on adding sensors. In practice, adjusting network structures sometimes consumes less costs than adding sensors, thus these two schemes can be accordingly chosen by the distinct situations. Moreover, this procedure is able to minimize the number of pinned nodes.
\end{itemize}

The remainder of this paper is organized as follows. Section \ref{section-motivated.example} discusses structural observability of Boolean networks in details, which motivates the concept of SOGs and the polynomial-time criterion, together with \cite{Margaliot2019TAC2727}, in Section \ref{section-SOGs}. In order to obtain an SOG from an arbitrarily given marked digraph, Section \ref{section-minimum.realization} designs two polynomial-time algorithms to seek and mark (or control) the minimum number of vertices. In Section \ref{section-further.analysis}, the observability preservation of iteration (\ref{equ-DTS}) under sensor failure, strongly structural observability of finite-field networks, and the observer design of LSBNs are further analyzed. Finally, the simulation part considers the minimum realization problem of cactus graphs and the observers design of T-LGL survival signal networks with $29$ nodes.

{\bf Notations:} The following notations will be useful for concise expression throughout this paper, and the other symbols shall be introduced once they are utilized.
\begin{itemize}
  \item[(1)] $\mathscr{B}:=\{1,0\}$;
  \item[(2)] $\mathbb{R}$: the real number domain;
  \item[(3)] $\mathbb{R}^n$: the $n$-dimensional column vector over $\mathbb{R}$;
  \item[(4)] $\mathbb{N}$: the integer domain;
  \item[(5)] $[m,n]_{\mathbb{N}}:=\{m,m+1,\cdots,n\}$ with $m,n\in\mathbb{N}$ and $m<n$;
  \item[(6)] $\mid S \mid$: the cardinal number of set $S$;
  \item[(7)] $\mathscr{B}^{m \times n}$: the $(m \times n)$-dimensional Boolean matrices;
  \item[(8)] $e_{i,n}$: the $i$-th column of $n$-dimensional identity matrix $I_n$;
  \item[(9)] $A^\top$: the transposition of matrix $A$;
  \item[(10)] $\lfloor v_i \rfloor$: the index of $v_i$;
  \item[(10)] $\mathbb{F}_p$\footnote{In order to guarantee the existence of a multiplicative inverse, number $p$ is required to be prime for arbitrary finite field \cite{papantonopoulou2002algebra}.}: a set of finite elements $\{0,1,\cdots,p-1\}$, together with addition $+_p$ and multiplication $\times_p$, satisfying the following six properties: 1) closure; 2) commutativity; 3) associativity; 4) distributivity; 5) the existence of an additive identity and a multiplicative identity; 6) the existence of an additive inverse and a multiplicative inverse. Please refer to \cite{papantonopoulou2002algebra} for detailed introductions about finite fields;
  \item[(11)] $\mathbb{F}_p^{n \times n}$: the set of $(n \times n)$ matrices over finite field $\mathbb{F}_p$.
\end{itemize}

\section{A Motivated Example: Structural Observability of Boolean Networks}\label{section-motivated.example}
The primal example that inspires this paper is the structural observability of Boolean networks. Boolean networks can be regarded as a special kind of system (\ref{equ-DTS}), where every variable displays binary configurations (i.e., $1$ and $0$) \cite{kauffman1969jtb437}. Taking $h$ directly observable outputs into account, the standard mathematical model for Boolean networks is given in the same form of (\ref{equ-DTS}) as
\begin{equation}\label{equ-BN}
\left\{\begin{aligned}
&x_i[k+1]=f_i((x_j[k])_{j\in \mathbf{N}_i}),~i=1,2,\cdots,n \\
&y_p[k]=x_p[k],~p=1,2,\cdots,h
\end{aligned}\right.
\end{equation}
where variable $x_i[k]$ takes values in $\mathscr{B}$, and functions $f_i: \mathscr{B}^{\mid \mathbf{N}_i \mid} \rightarrow \mathscr{B}$, $i\in[1,n]_{\mathbb{N}}$, are {\em minimally} represented logical functions. Particularly, $x_p$, $p\in[1,h]_{\mathbb{N}}$, are termed as directly observable state variables, since their real-time states are observable by sensors.

Boolean network (\ref{equ-BN}) is well known as a useful type of models to describe various systems including genetic regulatory networks \cite{kauffman1969jtb437}, multi-agent systems \cite{fagiolini2013auto2339}, networked evolutionary games \cite{zhaogd2016IJAS2016} and so on. Until now, plentiful results on analysis and control of Boolean networks have emerged under their algebraic state space representation, such as stability/stabilization \cite{valcher2015aut21,guoyq2020tac,lir2013ieeetac1853,zhongjie2019new}, controllability \cite{margaliot2012aut1218,mengmin2019auto70}, observability \cite{zhangkz2020tac}, optimal control \cite{wuyh2018tac262,valcher2013tac1258,margaliot2013siam2869}, and other related problems \cite{Yuyongyuan2019TAC,liht2020siam3632}. However, these results tacitly approve that node dynamics $f_i$, $i\in[1,h]_{\mathbb{N}}$, are fully available but, as indeed, are difficult to be identified \cite{Azuma2015TCNS179,Azuma2019TCNS464,azuma2019tcbb}. In what follows, we would like to study the observability of Boolean networks from the perspective of network structure.

As usual, an observable Boolean network (\ref{equ-BN}) is defined as follows:
\begin{definition}
Boolean network (\ref{equ-BN}) is said to be observable if the output trajectories starting from distinct initial states $x_0\in \mathscr{B}^n$ and $\hat{x}_0\in \mathscr{B}^n$ at initial time $k=0$, denoted by $y[k;x_0]$ and $y[k;\hat{x}_0]$, will be distinguishable at certain time instant.
\end{definition}

Then, we provide some necessary acknowledgement on the standard graph theory. For an arbitrarily digraph $\mathrm{G}:=(\mathrm{V},\mathrm{E})$, a vertex sequence $\langle v_{i_0}v_{i_1}\cdots v_{i_s}\rangle$ is a walk in digraph $\mathrm{G}$ from vertex $v_{i_0}$ to $v_{i_s}$ if $(v_{i_q}, v_{i_{q+1}})\in \mathrm{E}$ holds for $q\in[0,s-1]_{\mathbb{N}}$. A walk $\langle v_{i_0}v_{i_1}\cdots v_{i_s} \rangle$ is a path with no repeated vertex. A cycle is a path $\langle v_{i_0}v_{i_1}\cdots v_{i_s}\rangle$ satisfying $i_0=i_s$.

Boolean network (\ref{equ-BN}) can be formally represented by a triple $B(\mathrm{G},f,[1,h]_{\mathbb{N}})$, where symbol $\mathrm{G}$ captures its dependency graph, $f:=(f_1,f_2,\cdots,f_n)$ stands for dynamics component, and $[1,h]_{\mathbb{N}}$ is the index set of directly observable vertices.

To proceed, the structural equivalence of Boolean networks (see, e.g., \cite{Azuma2015TCNS179,Azuma2019TCNS464,azuma2019tcbb}) is presented except that we require that the first $h$ vertices are always directly observable. We say two digraphs are identical, i.e., $\mathrm{G}_1=\mathrm{G}_2$, for $\mathrm{G}_1:=(\mathrm{V}_1,\mathrm{E}_1)$ and $\mathrm{G}_2:=(\mathrm{V}_2,\mathrm{E}_2)$, if $\mathrm{V}_1=\mathrm{V}_2$ and $\mathrm{E}_1=\mathrm{E}_2$.
\begin{definition}\label{def-equivalence}
Boolean network $B(\hat{\mathrm{G}},\hat{f},[1,h]_{\mathbb{N}})$ is said to be structurally equivalent to Boolean network $B(\mathrm{G},f,[1,h]_{\mathbb{N}})$ if $\hat{\mathrm{G}}=\mathrm{G}$ holds for their dependency graphs.
\end{definition}

\begin{definition}\label{def-structuralobservability}
Boolean network $B(\mathrm{G},f,[1,h]_{\mathbb{N}})$ is said to be structurally observable if its structurally equivalent Boolean networks are all observable.
\end{definition}

\begin{remark}
Although dynamics component $f$ appears in Definition \ref{def-structuralobservability}, it is worth noticing that arbitrary modifications in node dynamics $f$ would not affect observability only if it still remains the original dependency graph. Hence, we pay our attention to study interaction digraphs instead of full dynamics component from then onward. Besides, it also means that structural observability could efficiently cope with the inaccuracies to a certain extent in the process of modeling system.
\end{remark}

Then, we would like to establish a necessary and sufficient criterion for the structural observability of Boolean networks. To this end, the following preliminaries that were firstly presented in \cite{Margaliot2019TAC2727} and \cite{Margaliot2019IEEECSL210} are briefly presented to help the readers understand better. In the interaction digraph (or, dependency graph) of Boolean network (\ref{equ-BN}), the vertices corresponding to directly observable state nodes and non-directly observable state nodes are respectively called directly observable vertices and {\em simple} vertices.
\begin{definition}[see \cite{Margaliot2019TAC2727}]\label{definition-P1}
A simple vertex $v$ of digraph $\mathrm{G}$ is said to satisfy Property $P_1$ if there is a vertex $v'$ whose in-neighbor is unique as $v$. A digraph $\mathrm{G}$ is said to satisfy Property $P_1$ if all simple vertices in digraph $\mathrm{G}$ satisfy Property $P_1$.
\end{definition}

\begin{definition}[see \cite{Margaliot2019TAC2727}]\label{definition-P2}
A cycle $C$ that entirely consists of simple vertices is said to satisfy Property $P_2$ if there exists a vertex $v\not\in C$ such that certain vertex $v'\in C$ is the unique in-neighbor of vertex $v$. A digraph $\mathrm{G}$ is said to satisfy Property $P_2$ if each cycle that entirely consists of simple vertices satisfies Property $P_2$.
\end{definition}

In \cite{Margaliot2019TAC2727}, a conjunctive Boolean network has been proved to be observable if and only if its dependency graph satisfies both Properties $P_1$ and $P_2$. Besides, a Boolean network $B(\mathrm{G},f,[1,h]_{\mathbb{N}})$ is observable if its dependency graph satisfies both Properties $P_1$ and $P_2$ \cite{Margaliot2019IEEECSL210}. Thus, we can easily conclude the following theorem.
\begin{theorem}\label{theorem-stru.obser.}
A Boolean network $B(\mathrm{G},f,[1,h]_{\mathbb{N}})$ is structurally observable if and only if the corresponding dependency digraph $\mathrm{G}$ satisfies both Properties $P_1$ and $P_2$.
\end{theorem}

Or alternatively, Weiss and Margaliot proved that digraph $\mathrm{G}$ satisfying both Properties $P_1$ and $P_2$ if and only if it can be decomposed into several observed paths \cite{Margaliot2019TAC2727}.
\begin{definition}[see \cite{Margaliot2019TAC2727}]\label{definition-obser.path}
Path $P:=\langle{v}_{i_0}{v}_{i_1}\cdots {v}_{i_s}\rangle$ is said to be an observed path if the terminal vertex ${v}_{i_s}$ is the unique directly observable vertex on path $P$ and ${v}_{i_{k}}$ is the unique in-neighbor of ${v}_{i_{k+1}}$ for any $k\in[0,s-1]_{\mathbb{N}}$.
\end{definition}

Thus, the following corollary can be provided.
\begin{corollary}\label{corollary-stru.obser.}
A Boolean network $B(\mathrm{G},f,[1,h]_{\mathbb{N}})$ is structurally observable if and only if its dependency graph $\mathrm{G}$ can be decomposed into several observed paths.
\end{corollary}

\begin{remark}
The concept of structural observability is a new concept, which has not appeared before. Besides, we prove that the structural observability of Boolean networks can be well addressed by the techniques in \cite{Margaliot2019TAC2727} and \cite{Margaliot2019IEEECSL210}.
\end{remark}

\begin{remark}
As mentioned in Introduction, the criteria established on $2^n \times 2^n$ can only deal with the controllability of Boolean networks with $n \leq 25$, or the observability of Boolean networks with $n \leq 30$. By contract, structural observability only makes use of $n \times n$ interaction digraphs, and can be checked within polynomial time. Thus, this limitation on network size can be released.
\end{remark}

An illustrative example is given in Appendix I to facilitate readers understanding on structural observability for Boolean networks.

\section{Structurally Observable Graphs}\label{section-SOGs}
Motivated by Theorem \ref{theorem-stru.obser.}, we also would like to propose a type of digraphs that has the ability to ensure the observability of iteration (\ref{equ-DTS}) from the viewpoint of interaction digraphs. In what follows, this type of digraphs is termed as structurally observable graphs (SOGs). Actually, in this section, we would like to verify that a digraph satisfying both Properties $P_1$ and $P_2$ is exactly an SOG.

To consider the structural observability of iteration (\ref{equ-DTS}), we focus on the $[1,h]_{\mathbb{N}}$-marked digraph $\mathrm{G}$ that is obtained from interaction digraph by marking vertices $v_1,v_2,\cdots,v_h$ therein as directly observable vertices. In this setup, the positions of sensors in iteration (\ref{equ-DTS}) are also completely reflected on this $[1,h]_{\mathbb{N}}$-marked digraph $\mathrm{G}$.
\begin{definition}\label{def-SOG}
A $[1,h]_{\mathbb{N}}$-marked digraph $\mathrm{G}$ is said to be an SOG for iteration (\ref{equ-DTS}) if the corresponding interaction digraph being this $[1,h]_{\mathbb{N}}$-marked digraph $\mathrm{G}$ ensures that iteration (\ref{equ-DTS}) must be observable.
\end{definition}

In what follows, we will show that Properties $P_1$ and $P_2$ suffice to guarantee the observability of iteration (\ref{equ-DTS}), which is a more general system and can be used to describe many other systems as mentioned in Introduction.
\begin{assumption}\label{assumption-model-assumption}
Suppose that node dynamics $f_i$, $i\in[1,n]_{\mathbb{N}}$, are injective if they are single-variable.
\end{assumption}
\begin{remark}
Assumption \ref{assumption-model-assumption} obviously holds for linear iterations, Boolean networks and finite-field networks.
\end{remark}
\begin{theorem}\label{theorem-stru.obser.-DTDS}
Under Assumption \ref{assumption-model-assumption}, a $[1,h]_{\mathbb{N}}$-marked digraph $\mathrm{G}$ satisfying both Properties $P_1$ and $P_2$ is an SOG.
\end{theorem}
\begin{proof}
It has been known that a $[1,h]_{\mathbb{N}}$-marked digraph $\mathrm{G}$ that satisfies both Properties $P_1$ and $P_2$ can be decomposed into several observed paths $P^{obs}_1,P^{obs}_2,\cdots,P^{obs}_h$ (see, e.g., Proposition 1 in \cite{Margaliot2019TAC2727}). Besides, for distinct states $x,\tilde{x}\in\mathbb{R}^n$, there is a minimal integer $\iota\in[1,n]_{\mathbb{N}}$ such that their $\iota$-th components are mutually distinct, i.e., $x_\iota\neq\tilde{x}_\iota$. Fixing number $\iota$, vertex $v_\iota \in \mathrm{V}$ must lie on one of observed path $P^{obs}_{\tau(\iota)}$ with $\tau(\iota)\in [1,h]_{\mathbb{N}}$. Without loss of generality, we assume the part of observed path $P^{obs}_{\tau(\iota)}$ from vertex $v_{\iota}$ to directly observable vertex $v_{\tau(\iota)}$ to be $\langle v_{g_1}v_{g_2} \cdots v_{g_\tau}\rangle$ with $g_1:=l$ and $g_\tau:=\tau(\iota)$. Since vertex $v_{g_{i}}$ is the unique in-neighbor of vertex $v_{g_{i+1}}$ in the $[1,h]_{\mathbb{N}}$-marked digraph $\mathrm{G}$, we can calculate the measurement of sensor $y_{\tau(l)}$ as that $y_{\tau(l)}[\tau]=x_{g_1}[\tau]=f_{g_1} \circ f_{g_2} \circ \cdots \circ f_{g_\tau}(x_\iota)$\footnote{In this paper, we define the operation ``$\circ$" for two functions $f_1$ and $f_2$ as $f_1 \circ f_2 (x)=f_1(f_2(x))$.}. On account of that $x_\iota \neq \tilde{x}_\iota$ and $f_{g_1},f_{g_2}, \cdots, f_{g_\tau}$ are injective with respect to single variable, one can conclude that $y_{\tau(\iota)}[\tau; x_0] \neq y_{\tau(\iota)}[\tau; \tilde{x}_0]$ holds. It amounts to say that the output trajectories respectively starting from initial states $x$ and $\tilde{x}$ at time $k=0$ are distinguished at time instant $\tau$. Thus, the $[1,h]_{\mathbb{N}}$-marked digraph that satisfies both Properties $P_1$ and $P_2$ is an SOG.
\end{proof}

\begin{corollary}\label{corollary-stru.obser.-DTDS}
Under Assumption \ref{assumption-model-assumption}, a $[1,h]_{\mathbb{N}}$-marked digraph $\mathrm{G}$, which can be decomposed into several observed paths, is an SOG.
\end{corollary}

\begin{remark}
Digraphs with both Properties $P_1$ and $P_2$ were originally studied in \cite{Margaliot2019TAC2727}. One should note that Theorem \ref{theorem-stru.obser.-DTDS} and Corollary \ref{corollary-stru.obser.-DTDS} illustrate that this kind of digraphs can have more general applicability in fact, while studying the observability of a very general system (\ref{equ-DTS}). We accordingly formalize such digraphs as a feasible type of SOGs.
\end{remark}

\begin{remark}
Besides, the defined SOGs have the ability to study the observability problem of structural systems in the situations where interaction digraphs are {\em available} but node dynamics are {\em unknown}. Thus, SOGs also have some degree of robustness with respect to system model.
\end{remark}

\begin{remark}
It should be noticed that the SOGs in Definition \ref{def-SOG} are not only a feasible kind, but are also tight for some special classes of iteration (\ref{equ-DTS}). In fact, from Theorem \ref{theorem-stru.obser.-DTDS}, this kind of SOGs is necessary to guarantee the structural observability of Boolean networks. Thus, it is a tight one.
\end{remark}

\begin{example}
We provide two $\{1,2\}$-marked digraphs to illustrate the conditions of SOGs. Fig. \ref{fig-not-sog} is a seven-vertices digraph with two vertices being marked. Since vertex $v_6$ does not satisfy Property $P_1$, it is not an SOG. While for Fig. \ref{fig-is-sog}, it can be decomposed into three observed paths as $\langle v_1 \rangle$, $\langle v_4 v_5 v_2 \rangle$ and $\langle v_7 v_6 v_3 \rangle$, thus by Corollary \ref{corollary-stru.obser.-DTDS} it is an SOG.

\begin{figure}[htbp]
\centering
\subfigure[A digraph is not an SOG.]{
\includegraphics[scale=1.5]{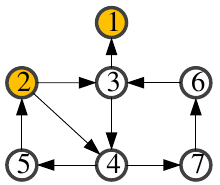}
\label{fig-not-sog}
}
\quad
\subfigure[A digraph is an SOG.]{
\includegraphics[scale=1.5]{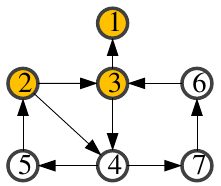}
\label{fig-is-sog}
}
\caption{Two $\{1,2\}$-marked digraphs are presented, where the left subgraph is not an SOG and the right subgraph is an SOG.}\label{fig-acyclic-I-II}
\end{figure}
\end{example}

\section{Minimum Realization Problems of Structurally Observable Graphs}\label{section-minimum.realization}
The attention of this section will be focused on how to mark or control the minimal vertices of an arbitrarily given digraph so as to derive a structurally observable graph (SOG).
\begin{problem}\label{problem-P1}
Given a non-marked digraph $\mathrm{G}$, how to mark the minimal number of vertices, denoted by $\mathrm{I}^\ast\subseteq \mathrm{V}$, so as to make the resulted digraph become an SOG?
\end{problem}

\begin{problem}\label{problem-P2}
Given a $[1,h]_{\mathbb{N}}$-marked digraph $\mathrm{G}$, how to modify the in-neighbors of the minimal number of vertices, denoted by $\Lambda^\ast\subseteq \mathrm{V}$, to make the resulted digraph be an SOG?
\end{problem}

Hereafter, we say that a vertex has been controlled if its in-neighbors have been modified. It is of theoretical interest and of practical significance to discuss the above two problems. In fact, Problem \ref{problem-P1} studies a problem of adding sensors as less as possible on iteration (\ref{equ-DTS}) so as to make it become structurally observable. Since Theorem \ref{theorem-stru.obser.-DTDS} and Corollary \ref{corollary-stru.obser.-DTDS} are same as the observability condition of conjunctive Boolean networks, this problem indeed has been studied by Algorithm 2 in \cite{Margaliot2019TAC2727}. On the other hand, Problem \ref{problem-P2} can be regarded as that of controlling agents as less as possible makes iteration (\ref{equ-DTS}) to realize the structural observability. While Problem \ref{problem-P1} admits us to impose sensors on agents, Problem \ref{problem-P2} is to modify the signal transform between agents, that is, to reject/accept the service of some other agents for the minimal number of agents. The following results will be displayed to give solutions to Problems \ref{problem-P1} and \ref{problem-P2} including the minimum vertex number and the corresponding positions. These results will help us to choose the optimal control strategy once we know the costs to add a sensor and to modify the in-neighbors of an agent.

\subsection{The Study of Problem \ref{problem-P1}}\label{subsection-P1}
In \cite{Margaliot2019TAC2727}, a polynomial-time algorithm has been developed to solve the minimal observability problem of conjunctive Boolean networks. Nonetheless, it can exactly tackle Problem \ref{problem-P1}, since Theorem \ref{theorem-stru.obser.-DTDS} has been established based on the same conditions as Theorem 1 in \cite{Margaliot2019TAC2727}. Thus, the polynomial-time Algorithm 1 in \cite{Margaliot2019TAC2727} could be directly utilized to solve Problem \ref{problem-P1}. We present Algorithm \ref{algorithm-minimumobservability} directly, and readers can refer to \cite{Margaliot2019TAC2727} for more details about this algorithm.

\begin{algorithm}
\caption{Solving Problem \ref{problem-P1} for a non-marked digraph.}\label{algorithm-minimumobservability}
\begin{algorithmic}[1]
\Require A digraph $\mathrm{G}:=(\mathrm{V},\mathrm{E})$
\Ensure A minimum vertex set $\mathrm{I}^\ast\subseteq \mathrm{V}$ to Problem \ref{problem-P1}
\State{Establish two $n$-bit lists $L_1$ and $L_2$}
\For{$v_i \in \mathrm{V}$}
\If{vertex $v_i$ does not satisfy Property $P_1$}
\State{$L^1_i \leftarrow 1$}
\State{$L^2_i \leftarrow 0$}
\Else
\State{$L^1_i \leftarrow 0$}
\State{$L^2_i \leftarrow 1$}
\EndIf
\EndFor
\State Search cycles in $\mathrm{G}$ with all componential vertices satisfying $L^2_i=1$ and put these cycles into set $L_C$
\For{$C\in L_C$}
\If{cycle $C$ satisfies Property $P_2$}
\State{$L_C\leftarrow L_C\backslash\{C\}$}
\EndIf
\EndFor
\State{$\mathrm{I}_1 \leftarrow \{v_i\mid L_1^i=1\}$}
\State{$\mathrm{I}_2 \leftarrow \{ v_i\in L_C \mid \forall v_j\in L_C, ~i \leq j \}$}
\State\Return $\mathrm{I}^\ast\leftarrow\mathrm{I}_1 \cup \mathrm{I}_2$
\end{algorithmic}
\end{algorithm}

As indeed, Algorithm \ref{algorithm-minimumobservability} is nothing different from Algorithm 2 in \cite{Margaliot2019TAC2727} except that we apply it in a more general iteration (\ref{equ-DTS}).

\begin{remark}
As formulated in \cite{Margaliot2019TAC2727}, the time complexity of Algorithm \ref{algorithm-minimumobservability} is $O(n^2)$, thus this approach can also be applied to SOGs that correspond to large-scale system (\ref{equ-DTS}).
\end{remark}

\subsection{The Study of Problem \ref{problem-P2}}\label{subsection-P2}
This subsection attempts to provide a polynomial-time algorithm to address Problem \ref{problem-P2} for an arbitrary $[1,h]_{\mathbb{N}}$-marked digraph $\mathrm{G}$. To this end, we respectively discuss it in two situations in the light of whether the given $[1,h]_{\mathbb{N}}$-marked digraph $\mathrm{G}$ containing cycles. That is, with regard to Problem \ref{problem-P2}, we firstly consider case (1) where $[1,h]_{\mathbb{N}}$-marked digraph $\mathrm{G}$ is acyclic, and then extend the results to be appropriate for an arbitrarily given digraph $\mathrm{G}$ that may contain some cycles (i.e., case (2)). Since we will not add sensors anymore, it is necessary to assume that $h>0$. For an arbitrary $[1,h]_{\mathbb{N}}$-marked digraph $\mathrm{G}$, we denote by $\mathrm{N}_{\mathrm{G}}^\ast$ the minimal number of vertices that need to be controlled in Problem \ref{problem-P2}, that is, $\mathrm{N}^\ast_{\mathrm{G}}=\mid \Lambda^\ast \mid$. The result will be $\mathrm{N}_{\mathrm{G}}^\ast=+\infty$ if $h=0$ trivially.

\subsubsection{The case of an arbitrary acyclic $[1,h]_{\mathbb{N}}$-marked digraph}\label{subsection-acyclic-P2}
In $[1,h]_{\mathbb{N}}$-marked digraph $\mathrm{G}$, all vertices $v_g$, with $g\not\in[1,h]_{\mathbb{N}}$ and an out-neighbor $v_u\in\mathrm{V}$ such that $(v_k,v_u)\in \mathrm{E}$ can imply $k=g$, are put into set $M_{\mathrm{G}}$. Take its complementary set as $\bar{M}_{\mathrm{G}}:=\mathrm{V}\backslash M_{\mathrm{G}}$. Obviously, $M_{\mathrm{G}}$ and $\bar{M}_{\mathrm{G}}$ are respectively the sets of vertices that satisfy Property $P_1$ and do not. Furthermore, one can split the set $\bar{M}_{\mathrm{G}}$ into two disjoint parts as $\bar{M}_{\mathrm{G}}:=\bar{M}^1_{\mathrm{G}}\cup \bar{M}^2_{\mathrm{G}}$, where $\bar{M}^1_{\mathrm{G}}\subseteq \bar{M}_{\mathrm{G}}$ is the set of vertices with out-degree zero and $\bar{M}^2_{\mathrm{G}}:=\bar{M}_{\mathrm{G}}\backslash \bar{M}^1_{\mathrm{G}}$. Thus, it holds that $\bar{M}^1_{\mathrm{G}} \cap \bar{M}^2_{\mathrm{G}}=\emptyset$.

\begin{lemma}\label{theorem-mustcontrol}
Given an acyclic $[1,h]_{\mathbb{N}}$-marked digraph $\mathrm{G}$. In order to solve Problem \ref{problem-P2}, for any $v_k \in \bar{M}_{\mathrm{G}}$, there is a vertex $\phi_{v_k}\in \mathrm{V}$ that needs to be controlled so as to make vertex $v_k$ become its unique in-neighbor. Moreover, for any $v_1,v_2\in \bar{M}_{\mathrm{G}}$ with $v_1\neq v_2$, it holds that $\phi_{v_1}\neq \phi_{v_2}$.
\end{lemma}
\begin{proof}
Denote by $\hat{\mathrm{G}}:=(\hat{\mathrm{V}},\hat{\mathrm{E}})$ the resulted SOG that is derived by controlling the vertices in $\Lambda^\ast$ of the given acyclic $[1,h]_{\mathbb{N}}$-marked digraph $\mathrm{G}$. To verify the former part of this theorem by seeking a contradiction, if there is a vertex $v_k \in \bar{M}_{\mathrm{G}}$ for which $(v_k,v_e)\not\in \hat{\mathrm{E}}$ holds for any $v_e\in\Lambda^\ast$, digraph $\hat{\mathrm{G}}$ obviously will not be an SOG since vertex $v_k$ does not satisfy Property $P_1$. It implies the satisfactory of former part.

Besides, we suppose that $\phi_{v_1}=\phi_{v_2}:=\phi_{v}$ holds for two distinct simple vertices $v_1$ and $v_2$, and seek a contradiction. For this case, one has that $(v_1,\phi_v)\in \hat{\mathrm{E}}$ and $(v_2,\phi_v)\in \hat{\mathrm{E}}$. Besides, since digraph $\hat{\mathrm{G}}$ is an SOG, it claims that vertices $v_1$ and $v_2$ are respectively the unique in-neighbor of two other vertices $v_{k_1}$ and $v_{k_2}$. Thus, removing vertex $\phi_v$ from the set $\Lambda^\ast$ would not affect the satisfactory of Property $P_1$ for SOG $\hat{\mathrm{G}}$. It contradicts with the minimality of set $\Lambda^\ast$, and the proof of this theorem is completed.
\end{proof}

\begin{remark}
Noticing that $\phi_v$ is a one-by-one correspondence, thus its inverse mapping $\phi^{-}_v$ can be well defined.
\end{remark}

In fact, Theorem \ref{theorem-mustcontrol} has shown that there exists at least a distinct vertex $\phi_{v_k}$ that needs to be controlled for different vertex $v_k\in \bar{M}_{\mathrm{G}}$, thus the following corollary can be established.
\begin{corollary}\label{corollary-geq}
Consider an arbitrarily given acyclic $[1,h]_{\mathbb{N}}$-marked digraph $\mathrm{G}$. It holds that $\mathrm{N}^\ast_{\mathrm{G}}\geq \mid \bar{M}_{\mathrm{G}} \mid$.
\end{corollary}

%
%

In the following, we will verify the main result to Problem \ref{problem-P2} for an acyclic $[1,h]_{\mathbb{N}}$-marked digraph $\mathrm{G}$. Before that, some necessary preliminaries for the proof of this theorem are listed:
\begin{itemize}
  \item[$\ast$] For a given digraph $\mathrm{G}:=(\mathrm{V},\mathrm{E})$, its corresponding adjacency matrix $A_{\mathrm{G}}:=(a_{ij})_{n \times n}$ can be established as that $a_{ij}=1$ if $(v_i,v_j)\in\mathrm{E}$, and $a_{ij}=0$ otherwise.
  \item[$\ast$] Distance $d_{i \rhd j}>0$ from vertex $v_i$ to vertex $v_j$ in digraph $\mathrm{G}$ is defined as the minimal integer $\kappa$ satisfying $[A^\kappa_{\mathrm{G}}]_{ij}=1$. The distance $d_{S \rhd j}>0$ from a set $S$ to vertex $v_j$ is computed as $d_{S \rhd j}:=\min\limits_{v_i \in S} d_{i \rhd j}$.
  \item[$\ast$] A undirected graph $\tilde{\mathrm{G}}=(\tilde{\mathrm{V}},\tilde{\mathrm{E}})$ is said to be a bipartite graph if its vertices can be colored by two colors while ensuring that there does not exist an edge $(v_i,v_j)\in \tilde{\mathrm{E}}$ with $v_i$ and $v_j$ being the same color.
  \item[$\ast$] For a given undirected graph $\tilde{\mathrm{G}}:=(\tilde{\mathrm{V}},\tilde{\mathrm{E}})$, a subset $\chi_M(\tilde{\mathrm{E}})$ of edge set $\tilde{\mathrm{E}}$ is called its matching if every vertex in $\tilde{\mathrm{V}}$ connects with one edge in $\tilde{\mathrm{E}}$ at most. Subset $\chi^\ast_M(\tilde{\mathrm{E}})$ is said to be the maximum matching of graph $\tilde{\mathrm{G}}$, if $\mid\chi^\ast_M(\tilde{\mathrm{E}})\mid \geq \chi_M(\tilde{\mathrm{E}})$ holds for arbitrary matching $\chi_M(\tilde{\mathrm{E}})$ of graph $\tilde{\mathrm{G}}$. The subgraph induced by edge set $\chi^\ast_{M}(\tilde{\mathrm{E}})$ is denoted by $\chi^\ast_M(\tilde{\mathrm{G}}):=(\chi^\ast_M(\tilde{\mathrm{V}}),\chi^\ast_M(\tilde{\mathrm{E}}))$.
  \item[$\ast$] The maximum matching of a bipartite graph $\tilde{\mathrm{G}}$ could be searched by calling Hopcroft-Krap algorithm within time $O(\sqrt{\mu}(\mu + \nu))$ \cite{hopcroft1973n}, where $\mu=\mid \tilde{\mathrm{V}} \mid$ and $\nu=\mid \tilde{\mathrm{E}} \mid$.
  \item[$\ast$] Path $P:\langle v_{i_s}\cdots v_{i_1}v_{i_0}\rangle$ is said to be an observed-path-compatible path (OP-CP) in the $[1,h]_{\mathbb{N}}$-marked digraph $\mathrm{G}$ if vertex $v_{i_k}$ is the unique in-neighbor of vertex $v_{i_{k-1}}$ for any $k\in[1,s]_{\mathbb{N}}$. Furthermore, path $P$ is called a Type I OP-CP if it is an OP-CP and the terminal vertex $v_{i_0}$ is marked; otherwise, OP-CP $P$ is called a Type II OP-CP.
\end{itemize}

\begin{theorem}\label{theorem-=}
Consider an arbitrarily given acyclic $[1,h]_{\mathbb{N}}$-marked digraph $\mathrm{G}$. It holds that $\mathrm{N}^\ast_{\mathrm{G}}=\mid \bar{M}_{\mathrm{G}} \mid$.
\end{theorem}
\begin{proof}
For the sake of clarity, we remove the detailed proof of this theorem into Appendix II.
\end{proof}

To conclude, based on the proof of Theorem \ref{theorem-=}, we present Algorithm \ref{algorithm-acyclic-P2} that determines these $\mid \bar{M}_{\mathrm{G}} \mid$ key vertices for an arbitrarily given acyclic $[1,h]_{\mathbb{N}}$-marked digraph $\mathrm{G}$ to solve Problem \ref{problem-P2}. As two sub-procedures, Algorithm \ref{algorithm-calculat.set} and Algorithm \ref{algorithm-construct.graph} are respectively given to find the vertex set $\bar{M}_{\mathrm{G}}$ and to construct the undirected digraph $\tilde{\mathrm{G}}$.
\begin{algorithm}\caption{Calculating sets $\bar{M}_\mathrm{G}$, $\bar{M}^1_\mathrm{G}$ and $\bar{M}^2_\mathrm{G}$.}\label{algorithm-calculat.set}
\begin{algorithmic}[1]
\Require A $[1,h]_{\mathbb{N}}$-marked digraph $\mathrm{G}=(\mathrm{V},\mathrm{E})$.
\Ensure Vertex sets $\bar{M}_\mathrm{G}$, $\bar{M}^1_\mathrm{G}$ and $\bar{M}^2_\mathrm{G}$
\For{$v_i\in\mathrm{V}$}
\If{out-degree of $v_i$ is zero}
\State{$\bar{M}^1_{\mathrm{G}}\leftarrow \bar{M}^1_{\mathrm{G}}\cup \{v_i\}$}
\If{$(v_i,v_j)\in \mathrm{E}$ and in-degree of $v_j$ is one}
\State{$M_{\mathrm{G}}\leftarrow M_{\mathrm{G}}\cup \{v_j\}$}
\EndIf
\EndIf
\EndFor
\State\Return{$\bar{M}_{\mathrm{G}}\leftarrow \mathrm{V}\backslash M_{\mathrm{G}}$, $\bar{M}^1_{\mathrm{G}}$ and $\bar{M}^2_{\mathrm{G}} \leftarrow \bar{M}_{\mathrm{G}} \backslash \bar{M}^1_{\mathrm{G}}$}
\end{algorithmic}
\end{algorithm}

\begin{algorithm}\caption{Constructing digraph $\tilde{\mathrm{G}}=(\tilde{\mathrm{V}},\tilde{\mathrm{E}})$.}\label{algorithm-construct.graph}
\begin{algorithmic}[1]
\Require A $[1,h]_{\mathbb{N}}$-marked digraph $\mathrm{G}=(\mathrm{V},\mathrm{E})$ and set $\bar{M}_{\mathrm{G}}$.
\Ensure The bipartite undirected graph $\tilde{\mathrm{G}}=(\tilde{\mathrm{V}},\tilde{\mathrm{E}})$
\For{$v_i\in\mathrm{V}$}
\If{$(v_i.v_j)\in\mathrm{E}$ and $v_i\in\bar{M}_{\mathrm{G}}$}
\State{$\nabla_1(\bar{M}_{\mathrm{G}}) \leftarrow \nabla_1(\bar{M}_{\mathrm{G}}) \cup \{v_j\}$}
\EndIf
\EndFor
\For{$v_j\in \nabla_1(\bar{M}_{\mathrm{G}}) \cap \bar{M}_{\mathrm{G}}$}
\State{$\mathrm{V}'\leftarrow\mathrm{V}'\cup\{v'_j\}$}
\EndFor
\For{$(v_i,v_j)\in\mathrm{E}$ with $v_i\in\bar{M}_{\mathrm{G}}$}
\If{$v_j\in \nabla_1(\bar{M}_{\mathrm{G}}) \backslash \bar{M}_{\mathrm{G}}$}
\State{$\tilde{\mathrm{E}} \leftarrow \tilde{\mathrm{E}}\cup (v_i,v_j)$}
\Else
\State{$\tilde{\mathrm{E}} \leftarrow \tilde{\mathrm{E}}\cup (v_i,v'_j)$}
\EndIf
\EndFor
\State\Return{bipartite undirected graph $\tilde{\mathrm{G}}=(\tilde{\mathrm{V}},\tilde{\mathrm{E}})$}
\end{algorithmic}
\end{algorithm}

\begin{algorithm}[h!]
\caption{Solving Problem \ref{problem-P2} for acyclic digraphs.}\label{algorithm-acyclic-P2}
\begin{algorithmic}[1]
\Require A $[1,h]_{\mathbb{N}}$-marked digraph $\mathrm{G}=(\mathrm{V},\mathrm{E})$.
\Ensure Vertex set $\Lambda^\ast$, number $\mathrm{N}_{\mathrm{G}}^\ast$ and function $\phi^{-}$
\State{call Algorithm \ref{algorithm-calculat.set} to print the sets $\bar{M}_{\mathrm{G}}$, $\bar{M}^1_{\mathrm{G}}$ and $\bar{M}^2_{\mathrm{G}}$}
\State{call Algorithm \ref{algorithm-construct.graph} to output a bipartite undirected graph $\tilde{\mathrm{G}}$}
\State{compute its maximum matching $\chi^\ast_M(\tilde{\mathrm{E}})$ and its induced subgraph $\chi_M^\ast(\tilde{\mathrm{G}})$}
\If{$\frac{\mid\chi^\ast_M(\tilde{\mathrm{V}})\mid}{2} < \mid\bar{M}_{\mathrm{G}}^2\mid$}
\State{$\bar{M}^1_{\mathrm{G}} \leftarrow \bar{M}^1_{\mathrm{G}}\cup\{\bar{M}^2_{\mathrm{G}}\backslash \chi^\ast_M(\tilde{\mathrm{V}})\}$}
\State{$\bar{M}^2_{\mathrm{G}} \leftarrow \bar{M}^2_{\mathrm{G}}\cup \chi^\ast_M(\tilde{\mathrm{V}})$}
\State{$\mathrm{E}\leftarrow \mathrm{E}\backslash \{(v_r,v_c)\in\mathrm{E}\mid v_r\in\bar{M}^2_{\mathrm{G}}\backslash \chi^\ast_M(\tilde{\mathrm{V}})\}$}
\State{$\mathrm{G}\leftarrow (\mathrm{V},\mathrm{E})$}
\EndIf
\State{$\Lambda^\ast \leftarrow ((\nabla_1(\bar{M}_{\mathrm{G}})\backslash\bar{M}_{\mathrm{G}}) \cap \chi^\ast_M(\tilde{\mathrm{V}})) \cup \{ v_j \mid v_j'\in \chi^\ast_M(\tilde{\mathrm{V}}) \cap \mathrm{V}'\}$}
\For{$v_j\in(\nabla_1(\bar{M}_{\mathrm{G}})\backslash\bar{M}_{\mathrm{G}}) \cap \chi^\ast_M(\tilde{\mathrm{V}})$}
\For{$(v_i,v_j) \in \chi^\ast_M(\tilde{\mathrm{E}})$}
\State{$\phi^{-}_{v_j} \leftarrow v_i$}
\State{$In.(v_j)\leftarrow\{v_i\}$}
\EndFor\EndFor
\For{$v_j\in\{v_j \mid v_j'\in \chi^\ast_M(\tilde{\mathrm{V}})\cap\mathrm{V}'\}$}
\For{$(v_i,v'_j) \in \chi^\ast_M(\tilde{\mathrm{E}})$}
\State{$\phi^{-}_{v_j} \leftarrow v_i$}
\State{$In.(v_j)\leftarrow\{v_i\}$}
\EndFor\EndFor
\If{$\bar{M}^1_{\mathrm{G}}\neq\emptyset$}
\State{Search Type I OP-CPs $P^{obs}_{\mathrm{I},1},P^{obs}_{\mathrm{I},2},\cdots, P^{obs}_{\mathrm{I},\alpha}$ and Type II OP-CPs $P^{obs}_{\mathrm{II},1},P^{obs}_{\mathrm{II},2},\cdots, P^{obs}_{\mathrm{II},\beta}$ in digraph $\mathrm{G}$}
\State{$\Lambda^\ast \leftarrow \Lambda^\ast \cup \left(\bigcup_{i=1}^{\beta-1}head.(P^{obs}_{\mathrm{II},i})\right)$}
\State{$\phi^{-}_{head.(P^{obs}_{\mathrm{I},\alpha})} \leftarrow tail.(P^{obs}_{\mathrm{II},1})$}
\For{$v_j= head.(P^{obs}_{\mathrm{II},k})$, $i\in[1,\beta-1]_{\mathbb{N}}$}
\State{$v_i \leftarrow tail.(P^{obs}_{\mathrm{II},k+1})$}
\EndFor
\EndIf
\State\Return{$\Lambda^\ast$, $\mathrm{N}^\ast_{\mathrm{G}}$ and $\phi^{-}$}
\end{algorithmic}
\end{algorithm}

Finally, we analyze time complexity of Algorithm \ref{algorithm-acyclic-P2}.  Once system (\ref{equ-DTS}) has been presented, the time complexity to construct the corresponding interaction digraph will be linear with vertex number and edge number, that is, $O(n + m)$, where $n:=\mid \mathrm{V} \mid$ and $m:=\mid \mathrm{E} \mid$. The complexity of implementing Algorithm \ref{algorithm-calculat.set} and Algorithm \ref{algorithm-construct.graph} is respectively $O(m)$ and $O(n+m)$. Expect for lines $3$ and $24$, the complexity of other steps is linear with either $n$ or $m$, that is, either $O(n)$ or $O(m)$. Since one needs $O(\sqrt{n}(n+m))$ and $O(n+m)$ to search the maximum matching of digraph $\tilde{\mathrm{G}}$ and search OP-CPs or OP-CCs respectively, the total time complexity of Algorithm \ref{algorithm-acyclic-P2} is $O((\sqrt{n}+1)(n+m))$.

\subsubsection{The case of arbitrary $[1,h]_{\mathbb{N}}$-marked digraph that contains cycles}
As for an arbitrarily given $[1,h]_{\mathbb{N}}$-marked digraph $\mathrm{G}$ that contains some cycles, the result would be more complex than that for the acyclic digraphs. To begin with, we define the observed-path-compatible cycles (OP-CCs). In what follows, we only consider the OP-CCs in the given $[1,h]_{\mathbb{N}}$-marked digraph $\mathrm{G}$, since the satisfactory of Property $P_1$ for vertices in cycle $C$, which is not an OP-CC, implies that cycle $C$ must satisfy Property $P_2$.
\begin{definition}
Cycle $C=\langle v_{i_1}v_{i_2},\cdots,v_{i_s} \rangle$ is called an OP-CC if vertex $v_{i_{j}}$, $j\in[1,s-1]_{\mathbb{N}}$, is the only in-neighbor of $v_{i_{j+1}}$ and vertex $v_{i_s}$ is the only in-neighbor of $v_{i_1}$.
\end{definition}

For a given $[1,h]_{\mathbb{N}}$-marked digraph $\mathrm{G}$, we could collect all its OP-CCs by set $S_{\mathrm{G}}$. Once the set $S_{\mathrm{G}}$ has been obtained, one can check whether OP-CC $C \in S_{\mathrm{G}}$ satisfies Property $P_2$. Therefore, one can split set $S_{\mathrm{G}}$ into two subsets $S^1_{\mathrm{G}}$ and $S^2_{\mathrm{G}}$, i.e., $S_{\mathrm{G}}:=S^1_{\mathrm{G}} \cup S^2_{\mathrm{G}}$, where $S^1_{\mathrm{G}}$ and $S^2_{\mathrm{G}}$ are respectively the sets of OP-CCs that satisfy Property $P_2$ and do not.

Then, a result whose role is same as Corollary \ref{corollary-geq} is given.
\begin{theorem}\label{theorem-cyclic->}
Consider an arbitrarily given $[1,h]_{\mathbb{N}}$-marked digraph $\mathrm{G}$. It holds that $\mathrm{N}^\ast_{\mathrm{G}}\geq \mid \bar{M}_{\mathrm{G}} \mid + \mid S^2_{\mathrm{G}} \mid$.
\end{theorem}
\begin{proof}
By resorting to Theorem \ref{theorem-mustcontrol}, one can similarly prove that, for the vertices in $\bar{M}_{\mathrm{G}}$, we must find $\mid \bar{M}_{\mathrm{G}} \mid$ distinct vertices and control them. Denote these $\mid \bar{M}_{\mathrm{G}} \mid$ controlled vertices as $\{v_{\alpha_1}, v_{\alpha_2}, \cdots, v_{\alpha_{\mid \bar{M}_{\mathrm{G}} \mid}}\}$ without loss of generality. Then, for every cycle $C\in S^2_{\mathrm{G}}$, we must control another distinct vertex $v_C \in C$ such that certain vertex $\beta_C$ becomes the unique in-neighbor of vertex $v_C$. Otherwise, the controlled digraph $\hat{\mathrm{G}}$ cannot satisfy Property $P_2$.

Moreover, according to the proof of Theorem \ref{theorem-mustcontrol}, vertices $\beta_C$ are mutually distinct for different OP-CCs and also do not include in $\{v_{\alpha_1}, v_{\alpha_2}, \cdots, v_{\alpha_{\mid \bar{M}_{\mathrm{G}} \mid}}\}$. Thus, we only need to prove the situation where certain vertex of $\{v_{\alpha_1}, v_{\alpha_2}, \cdots, v_{\alpha_{\mid \bar{M}_{\mathrm{G}} \mid}}\}$ locates in an OP-CC $C$, that is, $v_{\alpha_j}\in C$ for certain $j\in[1,\mid \bar{M}_{\mathrm{G}} \mid]_{\mathbb{N}}$. It could be described in Fig. \ref{fig-cyclic-disjoint}. If such case happens, that is, $\phi_{v_g}=v_k$, because cycle $C$ is an OP-CC, vertex $v_h$ would not satisfy Property $P_1$ once arc $(v_h,v_k)$ has been deleted from digraph $\mathrm{G}$. Thus, the removal of arc $(v_h,v_k)$ results in that vertex $v_h$ can be seemed as a new vertex in $\bar{M}^1_{\mathrm{G}}$ and we still need to find another vertex $\phi_{v_h}$ that needs to be controlled.
\begin{figure}[!ht]
\centering
\includegraphics[width=0.12\textwidth=0.5]{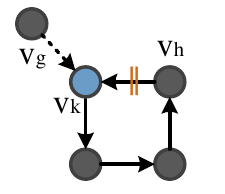}
\caption{This graph shows that situations where $v_{\alpha_j}\in C$ holds for some $j\in [1,\mid\bar{M}_{\mathrm{G}}\mid]_{\mathbb{N}}$ will not affect the number of controlled vertex for OP-CC $C$.}\label{fig-cyclic-disjoint}
\end{figure}

Therefore, one has that $\mathrm{N}^\ast_{\mathrm{G}}\geq \mid \bar{M}_{\mathrm{G}} \mid + \mid S^2_{\mathrm{G}} \mid$.
\end{proof}

The establishment of Theorem \ref{theorem-cyclic->} means that if we can control $\mid \bar{M}_{\mathrm{G}} \mid + \mid S^2_{\mathrm{G}} \mid$ vertices for an arbitrarily given $[1,h]_{\mathbb{N}}$-marked digraph $\mathrm{G}$ to make it become an SOG, then we have solved Problem \ref{problem-P2}. To this end, we control the digraph $\mathrm{G}$ via two steps. In the first step, we control $\mid \bar{M}_{\mathrm{G}} \mid$ vertices to make vertices in $\bar{M}_{\mathrm{G}}$ satisfy Property $P_1$. In the second step, we control the other $\mid S^2_{\mathrm{G}} \mid$ vertices to construct an SOG $\hat{\mathrm{G}}$.

\begin{figure}[!ht]
\centering
\includegraphics[width=0.45\textwidth=0.5]{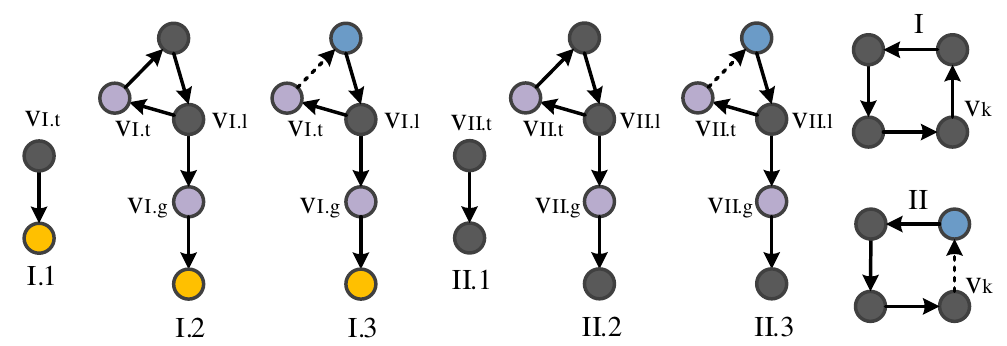}
\caption{The eight kinds of elementary components of digraph $\vec{\mathrm{G}}$ after the first step are provided, where the yellow vertices stand for the directly observable vertices, the purple vertices are the two out-neighbors of ``backtracking'' vertices, and the blue vertices are the vertices that have been controlled in the first step.}\label{fig-cyclic-I2-II4}
\end{figure}

After the first step, the obtained digraph is denoted by $\vec{\mathrm{G}}$. We can obtain some elementary components shown in Fig. \ref{fig-cyclic-I2-II4} and prove that they span the digraph $\vec{\mathrm{G}}$ in Lemma \ref{lemma-4types}. Different from the acyclic case discussed in Subsection \ref{subsection-acyclic-P2}, the Type I OP-CPs in this part can be further split into several types.

We first normalize these types of OP-CPs and OP-CCs that are presented in Fig. \ref{fig-cyclic-I2-II4}:
\begin{itemize}
  \item[$\ast$] Path $P$ is called a Type I.1 (resp., Type II.1) OP-CP if it is a Type I (resp., Type II) OP-CP and there is no vertex $v_{\mathrm{I},l}$ on path $P$ such that vertex $v_{\mathrm{I},l}$ is the unique in-neighbor of its terminal vertex $v_{\mathrm{I},t}$;
  \item[$\ast$] Path $P$ is called a Type I.2 (resp., Type II.2) OP-CP if it is a Type I (resp., Type II) OP-CP and there is a vertex $v_{\mathrm{I},l}$ on path $P$ such that vertex $v_{\mathrm{I},l}$ is the unique in-neighbor of its terminal vertex $v_{\mathrm{I},t}$ but there is no vertices that have been controlled on the path from vertex $v_{\mathrm{I},t}$ to vertex $v_{\mathrm{I},l}$;
  \item[$\ast$] Path $P$ is called a Type I.3 (resp., Type II.3) OP-CP if it is a Type I (resp., Type II) OP-CP and there is a vertex $v_{\mathrm{I},l}$ on path $P$ such that vertex $v_{\mathrm{I},l}$ is the unique in-neighbor of its terminal vertex $v_{\mathrm{I},t}$ but there are some vertices that have been controlled on the path from vertex $v_{\mathrm{I},t}$ to $v_{\mathrm{I},l}$;
  \item[$\ast$] OP-CC $C$ is called a Type I OP-CC if OP-CC $C$ does not contain any vertex that has been controlled;
  \item[$\ast$] OP-CC $C$ is called a Type II OP-CC if OP-CC $C$ contains some vertices that have been controlled.
\end{itemize}

Subsequently, some lemmas are provided to facilitate the investigation. As for the Type I.2, Type I.3, Type II.2 and Type II.3 OP-CPs (see, e.g., Fig. \ref{fig-cyclic-I2-II4}), we know that vertex $v_{\mathrm{I},l}$ (resp. $v_{\mathrm{II},l}$) has two out-neighbors $v_{\mathrm{I},t}$ and $v_{\mathrm{I},g}$ (resp., $v_{\mathrm{II},t}$ and $v_{\mathrm{II},g}$), which have the unique in-neighbor $v_{\mathrm{I},l}$. We say such vertex $v_{\mathrm{I},l}$ is a ``backtracking'' vertex. Similarly, vertex $v_{\mathrm{II},l}$ is also a backtracking vertex. It is obvious that the backtracking vertex on each Type I.2, Type I.3, Type II.2 and Type II.3 OP-CP is unique according to the definition of OP-CPs.
\begin{lemma}
As for digraph $\vec{\mathrm{G}}$, vertices $v_{\mathrm{I,t}}$ and $v_{\mathrm{I,g}}$ (resp., vertices $v_{\mathrm{II,t}}$ and $v_{\mathrm{II,g}}$) whose in-neighbors are unique as the backtracking vertex cannot have been controlled.
\end{lemma}
\begin{proof}
Without loss of generality, we suppose that vertex $v_{\text{I,t}}$ (see, e.g., Fig. \ref{fig-cyclic-I2-II4}) has been controlled in the first step, and seek a contradiction via two cases.

\begin{itemize}
  \item[(1)] For the first case, vertex $v_{\text{I,g}}$ has not been controlled, then one can conclude that $v_{\text{I,l}}\not\in \bar{M}_{\mathrm{G}}$. Hence, vertex $v_{\mathrm{I,l}}$ does not belong to the definition domain of $\phi$. Hence, one can imply that $(v_{\mathrm{I},l},v_{\mathrm{I},t})\not\in\vec{\mathrm{E}}$ if vertex $v_{\mathrm{I},t}$ has been controlled;
  \item[(2)] Considering the second case where vertex $v_{\text{I,g}}$ has been controlled, then vertex $v_{\text{I,l}}$ has two out-neighbors that have been controlled in digraph $\vec{\mathrm{G}}$. It would lead to at least $\mid \bar{M}_{\mathrm{G}} \mid+1$ controlled vertices in the first step. Obviously, it is a contradiction as the designed procedure.
\end{itemize}

Thus, the vertices having a backtracking vertex as the unique in-neighbor cannot have been controlled.
\end{proof}

\begin{lemma}\label{lemma-terminal.vertex}
The terminal vertices on the Type II OP-CPs in digraph $\vec{\mathrm{G}}$ must satisfy Property $P_1$ in the original digraph $\mathrm{G}$, that is, do not belong to set $\bar{M}_{\mathrm{G}}$.
\end{lemma}
\begin{proof}
The correctness of this lemma can be easily established, otherwise we would like to find a controlled vertex $\phi_v$ such that $(v,\phi_v) \in \vec{\mathrm{E}}$. Thus, vertex $v$ cannot be the terminal vertex of certain OP-CP in digraph $\vec{\mathrm{G}}$.
\end{proof}

\begin{figure}[!ht]
\centering
\includegraphics[width=0.35\textwidth=0.5]{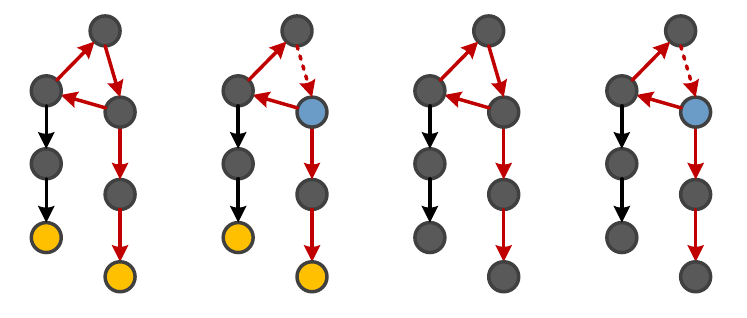}
\caption{The other types of components can be seemed as the combination of several types of basic components given in Fig. \ref{fig-cyclic-I2-II4}.}\label{fig-cyclic-decomposition}
\end{figure}

\begin{lemma}\label{lemma-4types}
The digraph $\vec{\mathrm{G}}$ can be spanned by Type I.1, I.2, I.3 OP-CPs, Type II.1, II.2, II.3 OP-CPs, as well as Type I and II OP-CCs (see, e.g., Fig. \ref{fig-cyclic-I2-II4}).
\end{lemma}
\begin{proof}
The last situation that is not included is presented in Fig. \ref{fig-cyclic-decomposition}, that is, there is an OP-CC that lies at the head of more than two Type I OP-CPs. Without loss of generality, we suppose that this OP-CC lies on two Type I OP-CPs. Thus, we could decompose it into a Type I.1 OP-CP and a Type I.2 OP-CP or a Type I.1 OP-CP and a Type I.3 OP-CP as in Fig. \ref{fig-cyclic-decomposition}. The cases when an OP-CC lies at the head of more than two OP-CPs can be similarly considered.
\end{proof}

\begin{theorem}\label{theorem-cyclic-=}
Consider an arbitrarily given $[1,h]_{\mathbb{N}}$-labelled digraph $\mathrm{G}$ containing some cycles. It holds that $\mathrm{N}^\ast_{\mathrm{G}}=\mid \bar{M}_{\mathrm{G}} \mid + \mid S^2_{\mathrm{G}} \mid$.
\end{theorem}
\begin{proof}
If we can generate an SOG from the given $[1,h]_{\mathbb{N}}$-labelled digraph $\mathrm{G}$ via controlling $\mid \bar{M}_{\mathrm{G}} \mid + \mid S^2_{\mathrm{G}} \mid$ vertices, this theorem can be established on the basis of Lemma \ref{theorem-cyclic->}. Note that Lemma \ref{lemma-4types} has proved that, by controlling vertices $\phi_{v_i}$ for all $v_i\in \bar{M}_{\mathrm{G}}$, the obtained digraph $\vec{\mathrm{G}}$ must be spanned by Type I.1, Type I.2, Type I.3 OP-CPs, Type II.1, Type II.2, Type II.3 OP-CPs, as well as Type I, Type II OP-CCs. In fact, up to now, the selection of these $\mid \bar{M}_{\mathrm{G}} \mid$ vertices can be arbitrary just making sure that vertices in $\bar{M}_{\mathrm{G}}$ satisfy Property $P_1$.

Therefore, we only discuss the second step about how to find another $\mid S^2_{\mathrm{G}} \mid$ controlled vertices such that the controlled digraph $\hat{\mathrm{G}}$ satisfies both Properties $P_1$ and $P_2$.

We first discuss the combination of two classes of OP-CCs and Type I OP-CPs. The combination for OP-CCs and Type II OP-CPs can be similarly considered.
Noting that Type II OP-CCs in digraph $\vec{\mathrm{G}}$ do not belong to set $S^2_{\mathrm{G}}$, thus we could not have assigned new controlled vertices for Type II OP-CCs.
\begin{figure}[htbp]
\centering
\subfigure[One Type I.3 OP-CP and one Type I OP-CC.]{
\includegraphics[scale=1.0]{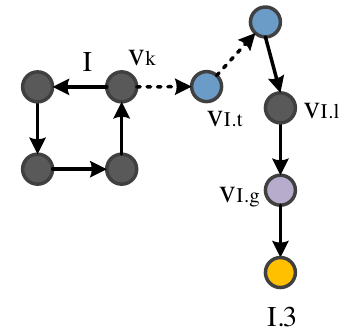}
\label{fig-cyclic-combine-I3-I1}
}
\quad
\subfigure[One Type I.3 OP-CP and one Type II OP-CC.]{
\includegraphics[scale=1.0]{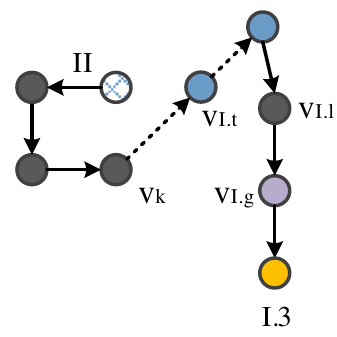}
\label{fig-cyclic-combine-I3-I2}
}
\quad
\subfigure[One Type II.3 OP-CP and one Type I OP-CC.]{
\includegraphics[scale=1.0]{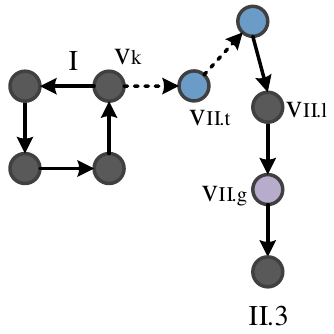}
\label{fig-cyclic-combine-II3-I}
}
\quad
\subfigure[One Type II.3 OP-CP and one Type II OP-CC.]{
\includegraphics[scale=1.0]{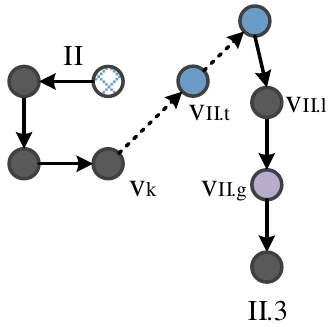}
\label{fig-cyclic-combine-II3-II}
}
\caption{The control for the combination of Type I (or, Type II) OP-CPs and Type I (or, Type II) OP-CCs is presented.}\label{fig-cyclic-combine}
\end{figure}

\begin{itemize}
  \item[$(1)$] As shown in Fig. \ref{fig-cyclic-combine-I3-I1}, to combine a Type I.3 OP-CP and Type I OP-CC, we could choose an arbitrary vertex $v_k$ in this OP-CC and make $\phi_{v_k}=v_{\mathrm{I},t}$. Then the vertices in this Type I OP-CC would satisfy Property $P_2$ and the resulted part after control is a Type I.2 OP-CP. This scene adds a controlled vertex.
  \item[$(2)$] As shown in Fig. \ref{fig-cyclic-combine-I3-I2}, to combine a Type I.3 OP-CP and Type II OP-CC, we could modify the original controlled vertex $\phi_{v_k}$ in this OP-CC as $v_{\mathrm{I},t}$, and the vertex $\phi_{v_k}$ will not be controlled anymore. Then, the combined part satisfies Property $P_1$ and can also be seemed to satisfy Property $P_2$ since this part does not contain a cyclic structure anymore. Consequently, this part will be a Type I.1 OP-CP. Since we just modify the original controlled vertex $\phi_{v_k}$ as $v_{\mathrm{I},t}$ in this case, it will not add the number of controlled vertex. It is feasible that because Type II OP-CC does not exist in original digraph $\vec{\mathrm{G}}$.
  \item[(3)] The control for other OP-CCs and OP-CPs can be implemented in the same manner as cases (1) and (2).
\end{itemize}

Finally, we only need to consider how to deal with the Type II OP-CPs. The reason why producing Type II OP-CPs rather than Type I OP-CPs are two situations, where one is that the terminal vertex of a Type II OP-CP lies in a Type I OP-CC, and the other is that the terminal vertex satisfies Property $P_1$ in the original digraph $\mathrm{G}$ and does not lie at a Type I OP-CC, but its out-neighbor has been controlled in the first step.
\begin{figure}[htbp]
\centering
\subfigure[The terminal vertex $v_{\mathrm{II},s}$ of Type II OP-CP lies in a Type I OP-CC.]{
\includegraphics[scale=1.0]{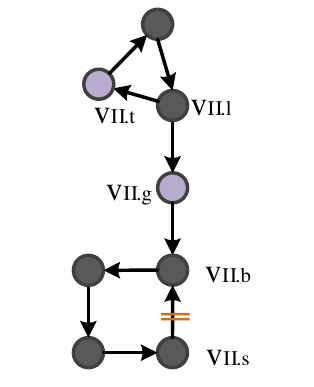}
\label{fig-cyclic-combine-typeii1}
}
\quad
\subfigure[the terminal vertex $v_{\mathrm{II},b}$ satisfies Property $P_1$ in the original digraph $\mathrm{G}$ but its out-neighbor $v_{\mathrm{II},s}$ has been controlled in the first step.]{
\includegraphics[scale=1.0]{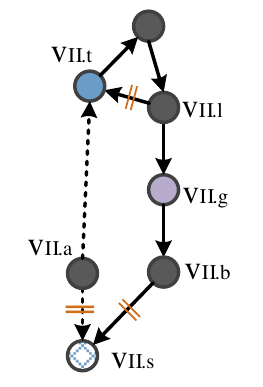}
\label{fig-cyclic-combine-typeii2}
}
\caption{The control for the terminal vertices of the Type II OP-CPs.}\label{fig-cyclic-combine-typeii}
\end{figure}

\begin{itemize}
  \item[(1)] Discuss the first case shown in Fig. \ref{fig-cyclic-combine-typeii1}. For this case, we need to choose a controlled vertex for this OP-CC in original digraph $\mathrm{G}$, but it is replaced by vertex $v_{\mathrm{II},s}$ that does not satisfy Property $P_1$ now. Thus, such OP-CP can be dealt with in the same manner as in Fig. \ref{fig-cyclic-combine} to combine with Type I OP-CPs.
  \item[(2)] For the second situation shown in Fig. \ref{fig-cyclic-combine-typeii2}, vertex $v_{\mathrm{II},s}$ is assumed to be the out-neighbor of vertex $v_{\mathrm{II},b}$ in the original digraph $\mathrm{G}$ and it has been controlled in digraph $\vec{\mathrm{G}}$. We change $\phi_{v_{\mathrm{II},a}}$ as vertex $v_{\mathrm{II},t}$ instead of the original vertex $v_{\mathrm{II},s}$. In this setting, we can obtain a new OP-CP without adding the number of controlled vertices.
\end{itemize}

In addition, it is obvious from Fig. \ref{fig-cyclic-combine-typeii1} and Fig. \ref{fig-cyclic-combine-typeii2} that the number of controlled vertices would plus one or remain unchanged for cases (1) and (2), respectively. Since Type I OP-CC originally exists in digraph $\mathrm{G}$, the remain controlled vertex number is $\mid S_{\mathrm{G}}^2 \mid$.

Therefore, the minimum number of controlled vertices is equal to $\mathrm{N}^\ast_{\mathrm{G}}=\mid \bar{M}_{\mathrm{G}} \mid + \mid S^2_{\mathrm{G}} \mid$.
\end{proof}

Consequently, we provide Algorithm \ref{algorithm-cyclic-P2} to control the minimal number of vertices to generate an SOG from an arbitrarily given digraph $\mathrm{G}$ that contains cycles.
\begin{algorithm}[h!]
\caption{Solving Problem \ref{problem-P2} for arbitrary digraphs.}\label{algorithm-cyclic-P2}
\begin{algorithmic}[1]
\Require A $[1,h]_{\mathbb{N}}$-marked digraph $\mathrm{G}=(\mathrm{V},\mathrm{E})$.
\Ensure Vertex set $\Lambda^\ast$ and function $\phi^{-}$
\State{call Algorithm \ref{algorithm-calculat.set} to print the set $\bar{M}_{\mathrm{G}}$}
\For{$v_i\in \bar{M}_{\mathrm{G}}$}
\State{let $\phi_{v_i}$ be an arbitrary vertex $v_j\in\mathrm{V}$}
\State{remove vertex $v_j$ from set $\mathrm{V}$}
\State{change the in-neighbor set of $v_i$ as $\phi_{v_i}^{-}$}
\EndFor
\State{search Type I, Type II OP-CPs and Type I, Type II OP-CCs in digraph $\mathrm{G}$}
\If{$\phi_{v_i}=v_j$ and $v_j$ lies in a Type I OP-CC $C_i$}
\If{OP-CC $C_i$ has a backtracking vertex $v'_i$}
\State{$\phi_{v_i}\leftarrow Out.(v_i')$ with $Out.(v_i') \in C_i$}
\Else
\State{let $\phi_{v_i}$ be an arbitrary vertex in $C_i$}
\EndIf
\EndIf
\If{there is a Type II OP-CP $P_i$ whose terminal vertex $v_i$ belong to neither $\bar{M}_{\mathrm{G}}$ nor an OP-CC}
\State{find $(\tilde{v},\phi_{v_i}) \in \mathrm{E}$}
\State{modify $\phi_{v_i}$ as $head.(P_i)$}
\State{let $\phi_{v_i}=\emptyset$}
\EndIf
\For{every Type II OP-CP}
\State{find its terminal vertex $v$}
\State{let $\phi_v$ be head vertex of another OP-CP}
\EndFor
\For{every Type II OP-CC}
\State{replace one of controlled vertices in this OP-CC as the head vertex of an OP-CP}
\EndFor
\For{every Type I OP-CC}
\State{choose an arbitrary vertex $v$ in it}
\State{let $\phi_v$ be the head vertex of an OP-CP}
\EndFor
\State\Return{function $\phi^{-}$ and the set $\Lambda^\ast$}
\end{algorithmic}
\end{algorithm}

Finally, we compute the time complexity of Algorithm \ref{algorithm-cyclic-P2}. The complexity of generating the interaction digraph is linear with vertex number $n$ and edge number $m$, i.e., $O(n+m)$. Additionally, except that the complexity of lines $15-19$ is bounded by $O(nm)$, the other parts are linear with vertex number $n$ or edge number $m$. Therefore, Algorithm \ref{algorithm-cyclic-P2} satisfies the bound $O(nm)$ for time complexity.

Considering the lack of cycles in Theorem \ref{theorem-=}, Theorem \ref{theorem-=} could be regraded as a special case of Theorem \ref{theorem-cyclic-=} where $\mid S^2_{\mathrm{G}} \mid=0$. Therefore, together with Theorem \ref{theorem-=} and Theorem \ref{theorem-cyclic-=}, we can provide the unifying solution to Problem \ref{problem-P2}, which is the most important result of this paper, as follows:
\begin{theorem}\label{theorem-all-=}
Consider an arbitrarily given $[1,h]_{\mathbb{N}}$-marked digraph $\mathrm{G}$. It holds that $\mathrm{N}^\ast_{\mathrm{G}}=\mid \bar{M}_{\mathrm{G}} \mid + \mid S^2_{\mathrm{G}} \mid$.
\end{theorem}

\begin{remark}
Since Algorithm \ref{algorithm-cyclic-P2} is proposed for marked digraphs that may contain cycles, it naturally suits for acyclic marked digraphs. However, it does not mean that Algorithm \ref{algorithm-acyclic-P2} is meaningless. Comparing their complexity $O((\sqrt{n}+1)(n+m))$ and $O(nm)$, if $m > \sqrt{n}+1$, then $O((\sqrt{n}+1)(n+m))<O(nm)$. Or alternatively, Algorithm \ref{algorithm-acyclic-P2} requires less time than Algorithm \ref{algorithm-cyclic-P2} to deal with an acyclic marked digraph satisfying $m > \sqrt{n}+1$. Hence, for a dense acyclic digraph, it is advised to utilize Algorithm \ref{algorithm-acyclic-P2} for the sake of reducing time cost.

\end{remark}

\section{Further Analysis of Structurally Observable Graphs}\label{section-further.analysis}
In this section, some further analysis of structurally observable graphs (SOGs) will be discussed, including the observability preservation of an SOG under the failure of a single sensor, the estimate of the minimal controlled vertices for randomly generated SOGs, the strongly structural observability of finite-field networks, as well as the observer design for Boolean networks via pinning control.

\subsection{Observability Preservation Under Sensor Failure}
In this subsection, we consider whether the observability property of an SOG can be preserved while the failure of a single sensor happens. The essential part of this problem is the position of fault sensor $y_\zeta$. According to Theorem \ref{theorem-=}, we know that the vital vertices for the SOGs can be divided into two parts--vertices that do not satisfy Property $P_1$ as well as vertices in OP-CCs that do not satisfy Property $P_2$. Thus, by Theorem \ref{theorem-stru.obser.-DTDS}, we can obtain the following theorem.
\begin{theorem}\label{theorem-sensor-failure}
The observability of a digraph cannot be preserved while sensor $y_\zeta$ fails if and only if one of following conditions holds:
\begin{itemize}
  \item[(1)] sensor $y_\zeta$ is imposed on a vertex that does not satisfy Property $P_1$; or
  \item[(2)] the OP-CC that sensor $y_\xi$ localizes at does not satisfy Property $P_2$ and sensor $y_\zeta$ is the unique sensor in this OP-CC.
\end{itemize}
\end{theorem}

\subsection{Estimation for Randomly Generated Labeled Digraphs}
This paper offers an alternative method to the approach of \cite{Margaliot2019TAC2727}. Given a $[1,h]_{\mathbb{N}}$-marked digraph and to generate an SOG, we know that $\mid \mathrm{I}^\ast \mid=\mid \Lambda^\ast \mid-h^\ast$ by observing Algorithm \ref{algorithm-minimumobservability} and Algorithm \ref{algorithm-cyclic-P2}, where the number of sensors that satisfy conditions (1) or (2) in Theorem \ref{theorem-sensor-failure} is denoted by $h^\ast$. Thus, one should choose a better method according to whether it is easier (with respect to cost or feasibility) to add sensors--and then should choose the approach in Subsection \ref{subsection-P1}, or easier to change the communication transmit between agents--and then should choose the method in Subsection \ref{subsection-P2}.

Compared with the optimal vertex number estimation $n^\ast$ for randomly (Erd\H{o}s-R\'{e}nyi) generated marked digraphs in \cite{Margaliot2019TAC2727}, we have $\mathrm{N}^\ast_{\mathrm{G}}=n^\ast-h^\ast$.

According to the estimation for $n^\ast$ in \cite{Margaliot2019TAC2727}, one can conclude that, at the optimal probability $p^\ast=n^{-1}$, the optimal topology structure admits (about) $(1-e^{-1})n-h^\ast\approx 0.632n-h^\ast$ vertices that need to be controlled in order to make a randomly generated digraph $\mathrm{G}$ become an SOG.

\subsection{Structural Observability of Finite-Field Networks}
In what follows, we consider the strongly structural observability of linear systems over finite field $\mathbb{F}_p$. On the basis of interaction digraph $\mathrm{G}$, we define a weighted adjacency matrix $A=(a_{ij})_{n \times n}$, $a_{ij}\in\mathbb{F}_p$, where $a_{ij}=0$ if $(v_i.v_j)\not\in\mathrm{E}$. Then, the evolution of network state over time is given as
\begin{equation}\label{equ-FFN}
x[k+1]=Ax[k]
\end{equation}
where $A$ is called the network matrix and all operations are performed over finite field $\mathbb{F}_p$. Before presenting the definition of strongly structural observability, the equivalent relationship ``$\sim$'' between two matrices is firstly given.
\begin{definition}
Given matrices $A\in\mathbb{F}_p^{n \times n}$ and $A'\in\mathbb{F}_p^{n \times n}$, we say $A \sim A'$ if the positions of positive elements in matrices $A$ and $A'$ are identical.
\end{definition}

Accordingly, the strongly structural observability of finite-field network (\ref{equ-FFN}) can be designed as follows:
\begin{definition}
Given finite-field network (\ref{equ-FFN}), it is said to be strongly structurally observable if, for any matrix $A'\in \mathbb{F}_p^{n \times n}$ satisfying that $A' \sim A$, any finite-field network $x[k+1]=A'x[k]$ must be observable.
\end{definition}

\begin{theorem}
Iteration (\ref{equ-FFN}) is strongly structurally observable if its interaction digraph satisfies both Properties $P_1$ and $P_2$.
\end{theorem}

Subsequently, we would like to design an observable finite-field network (\ref{equ-FFN}). To this end, an auxiliary digraph $\hat{\mathrm{G}}$ is constructed based on digraph $\mathrm{G}$. Let $\hat{\mathrm{V}}:=\mathrm{V}\cup\{v^-_1,v^-_2,\cdots,v^-_h\}$ and $\hat{\mathrm{E}}=(v_1,v_1^-)\cup(v_2,v_2^-)\cup\cdots\cup(v_h,v_h^-)\cup\mathrm{E}$, where vertex $v_i^{-}$ actually stands for output variable $y_i$.
\begin{theorem}
Finite-field network (\ref{equ-FFN}) is structurally observable if the minimum disjoint path cover problem for auxiliary digraph $\hat{\mathrm{G}}$ returns as $h$ paths cover.
\end{theorem}
\begin{proof}
If the solution to the minimum disjoint path cover problem of auxiliary digraph $\hat{\mathrm{G}}$ is $h$, since directly observable vertices must be the terminal vertices of these $h$ disjoint paths, then we can assign any non-zero weights on the oriented edges through these paths. Moreover, the weight $a_{ij}$ of other edges can be zero. As a consequence, the designed finite-field network will be observable by Theorem \ref{theorem-stru.obser.-DTDS}.
\end{proof}

\subsection{Pinning Observability of Boolean Networks}\label{subsection-bn-observer}
As illustrated in \cite{zhusy2020novel}, a group of feasible observers has been constructed for LSBNs via distributed pinning control strategy while node dynamics is available. The most essential part is to design a pinning control scheme so as to make the original network be observable. By means of Algorithm \ref{algorithm-cyclic-P2}, the minimum pinned node set with respect to interaction digraph is returned as $\Lambda^\ast$, in which the desired in-neighbor of each vertex $v\in\Lambda^\ast$ is denoted by $\phi_v^{-}$.

Once pinning controller, together with known pinned node set $\Lambda^\ast$, have been imposed on Boolean network (\ref{equ-BN}), the original Boolean network is turned into
\begin{equation}\label{equ-pinned-BN}
\left\{\begin{aligned}
&x_i[k+1]=u_i[k] \oplus_i f_i((x_j[k])_{j\in \mathbf{N}_i}),~i\in\Lambda^\ast \\
&x_i[k+1]=f_i((x_j[k])_{j\in \mathbf{N}_i}),~i\not\in\Lambda^\ast \\
&y_p[k]=x_p[k],~p=1,2,\cdots,h
\end{aligned}\right.
\end{equation}
the configurations that need to be designed are two parts, that is, control inputs $u_i[k]=g_i({\bf x}[k])$ and logical coupling $\oplus_i$, and the purpose is to control vertex $\phi^{-}_v$ such that it becomes the unique in-neighbor of vertex $v$, $\forall v\in\Lambda^\ast$.

To this end, variables $x_i[k]$ and $u_j[k]$ are respectively denoted into their corresponding canonical form ${\bm x}_i[k]$ and ${\bm u}_j[k]$, where ${\bm x}_i[k]:=({\bm x}_i[k],1-{\bm x}_i[k])^\top$ and ${\bm u}_i[k]:=({\bm u}_i[k],1-{\bm u}_i[k])^\top$. Let symbol ``$\ltimes$'' stands for the semi-tensor product of matrices\footnote{The semi-tensor product of matrices $A\in\mathbb{R}^{a \times b}$ and $B\in\mathbb{R}^{c \times d}$ is defined as $$A \ltimes B:=(A \otimes I_{t/b})(B \otimes I_{t/c})$$ where $t=\text{l.c.m.}\{b,c\}$ is the least common multiple of integers $b$ and $c$, and ``$\otimes$'' is the tensor product of matrices.} (see, e.g., monograph \cite{chengdz2011springer} for more details about semi-tensor product of matrices), the node dynamics of pinned nodes can be equivalently written as its algebraic form:
\begin{equation}\label{equ-pinned-BN-algebraic}
{\bm x}_i[k+1]=M_{\oplus_i}{\bm u}_i[k] M_{f_i} \left(\ltimes_{j\in\mathbf{N}_i}{\bm x}_j[k]\right),~i\in\Lambda^\ast
\end{equation}
where $M_{\oplus_i}$ and $M_{f_i}$ are the corresponding structure matrices\footnote{By denoting all its variables into canonical form, any given logical function $f$ can be equivalently converted into multi-linear form ${\bm f}$ as ${\bm f}=M_f \ltimes {\bm x}_1 \ltimes {\bm x}_2 \ltimes \cdots \ltimes {\bm x}_n$ where $M_f$ is called the structure matrix of $f$.} of logical functions $\oplus_i$ and $f_i$.

Overall, we can design the pinning control on each pinned node by three situations:
\begin{itemize}
  \item Type I pinned node $v_i$ (that is, ${\bf N}_i \neq \emptyset$ and $\lfloor\phi^{-}_{v_i}\rfloor \not\in {\bf N}_i$): Let state feedback input $u_i[k]=g_i((x_{j}[k])_{j\in{\bf N}_i\cup\{\lfloor\phi^{-}_{v_i}\rfloor\}})$. The desired structure matrices $M_{\oplus_i}$ and $M_{{\bm g}_i}$ can be solved by
  \begin{equation*}
\begin{aligned}
&M_{\oplus_i}M_{g_i}\left(I_{2^{k_i+1}}\otimes (M_{f_i}(I_{2^{\iota_i-1}}\otimes M_{d,2}))\right)M_{r,2^{k_i+1}}\\
&~~~~~~~~~~~~~~~~~~~~~~~~~~~~~~~~~~~~~=(\hat{A}_i\otimes {\bf 1}^\top_{2^{k_i}}) \mathrm{W}^\top_{\left[2,2^{\iota_i-1}\right]}
\end{aligned}
\end{equation*}
where $\hat{A}_i\in\{I_2,{\bf 1}_2 \otimes {\bf 1}_2^\top-I_2\}$, $k_i=\mid {\bf N}_i \mid$, number $\iota_i$ is the increasing order of $\lfloor\phi^{-}_{v_i}\rfloor$ in set ${\bf N}_i\cup\{\lfloor\phi^{-}_{v_i}\rfloor\}$, and $M_{d,2}$, $M_{r,2^{k_i+1}}$ and $\mathrm{W}_{\left[2,2^{\iota_i-1}\right]}$ are respectively the corresponding dimensional dummy matrix, power-reducing matrix, and swap matrix (see, e.g., \cite{chengdz2011ijrnc134} for more details).
  \item Type II pinned node $v_i$ (that is, ${\bf N}_i \neq \emptyset$ and $\lfloor\phi^{-}_{v_i}\rfloor \in {\bf N}_i$): Let state feedback input $u_i[k]=g_i((x_{j}[k])_{j\in{\bf N}_i})$. The desired structure matrices $M_{\oplus_i}$ and $M_{{\bm g}_i}$ can be calculated from
        \begin{equation*}
\begin{aligned}
M_{\oplus_i}M_{g_i}\left(I_{2^{k_i}}\otimes M_{f_i}\right)M_{r,2^{k_i}}=(\hat{A}_i\otimes {\bf 1}^\top_{2^{k_i-1}}) \mathrm{W}^\top_{\left[2,2^{\iota_i-1}\right]}
\end{aligned}
\end{equation*}
where $\iota_i$ is the increasing order of $\lfloor\phi^{-}_{v_i}\rfloor$ in set ${\bf N}_i$.
\item Type III pinned node $v_i$ (that is, ${\bf N}_i = \emptyset$): We can directly assign $g_i[k]=x_{\lfloor\phi_{v_i}\rfloor}[k]$ and $\oplus_i=\wedge$.
\end{itemize}

By above procedure, the minimum pinned nodes are selected by Algorithm \ref{algorithm-cyclic-P2}. Thus, the control cost in this paper would be lower than that in \cite{zhusy2020novel} from the viewpoint of controlled nodes. Once pinning controller has been designed, the observer can be immediately constructed as
\begin{equation}
\left\{\begin{aligned}
&\hat{x}_i[k+1]=g_i[k] \oplus_i f_i((\hat{x}_j[k])_{j\in \mathbf{N}_i}),~i\in\Lambda^\ast \\
&\hat{x}_i[k+1]=f_i((\hat{x}_j[k])_{j\in \mathbf{N}_i}),~i\not\in\Lambda^\ast
\end{aligned}\right.
\end{equation}
and the initial state can be estimated as
$$\left\{\begin{aligned} &x_i[0]= \hat{f}_{\xi^{\pi_i}_{\lambda_i}} \circ \hat{f}_{\xi^{\pi_i}_{\lambda_i-1}} \circ \cdots\circ \hat{f}_{\xi^{\pi_i}_{1}}(x_{\pi_i}[\lambda_i])\\ &\hat{f}_{i}[k]=\left\{\begin{aligned} & g_i[k] \oplus_i f_i[k],~i\in\Lambda^\ast \\ & f_i[k],~i\not\in\Lambda^\ast \end{aligned} \right.\end{aligned}\right.$$
where path $P:\langle\xi^{\pi_i}_{\lambda_i}\xi^{\pi_i}_{\lambda_i-1}\cdots\xi^{\pi_i}_{1}\rangle$ is the local part, from simple vertex $v_i$ to directly observable vertex $v_{\pi_i}$, of the localized observed path.

\begin{remark}
The time complexity of the existing observers construction for Boolean networks, based on exhaustive state space method, increases exponentially with the number of network nodes (see, e.g., \cite{Valcher2012TAC1390,zhangzh2020tac,zhangzh2019tcns516}), thus they are not suitable for LSBNs. By comparison, the above approach only requires the node-to-node information and $n \times n$ interaction digraph, thus the time complexity is bounded by $O(n2^K)$, where number $K$ is the largest in-degree of pinned nodes. It is worth noting that many biological networks are sparsely connected (see, e.g., \cite{jeong2000nature,jeong2001nature}), thus $K \ll n$ usually holds, which demonstrates the feasibility of this approach for LSBNs.
\end{remark}

\section{Applications to Cactus Graphs and Observer Design of Boolean Networks}
\subsection{Minimum Node Control For Cactus Graphs}
In this section, an important type of interaction digraphs--Cactus Graphs \cite{azuma2018tcns775cactus}--is discussed; it has useful network properties in many systems including diagonal stability \cite{arcak2011diagonal} as well as structural oscillatory of Boolean networks \cite{Azuma2019TCNS464}, and so on. Besides, the Cactus-Expandable Graphs were proposed and discussed by Azuma {\em et al.} in \cite{azuma2018tcns775cactus}.
\begin{definition}[see \cite{azuma2018tcns775cactus}]
A given digraph is called a cactus graph if it is connected and there is no edges contained in two or more than two distinct simple cycles.
\end{definition}

\begin{figure}[!ht]
\centering
\includegraphics[width=0.4\textwidth=0.5]{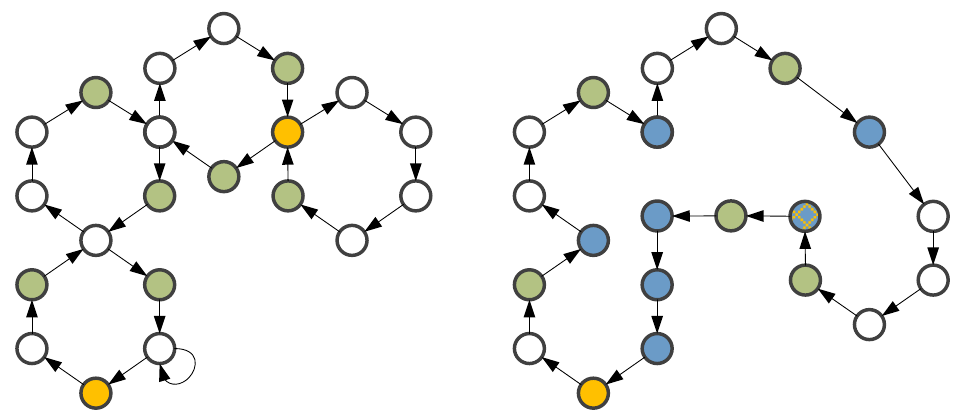}
\caption{A five-cycle-cactus is presented as left subgraph, where green and yellow vertices are respectively vertices that do not satisfy Property $P_1$ and directly observable vertices.}\label{fig-cactus-five-leaves}
\end{figure}

A cactus with five different cycles is presented as in left subgraph of Fig. \ref{fig-cactus-five-leaves}, where yellow vertices stand for the directly observable vertices. For this five-cycles-cactus, the number of vertices that do not satisfy Property $P_1$ is seven and colored by green. Besides, there is no OP-CCs thus, by Theorem \ref{theorem-all-=}, the minimal number of controlled vertices will be seven.

In above setting, set $\bar{M}_{\mathrm{G}}$ can be calculated as the seven green vertices in Fig. \ref{fig-cactus-five-leaves} and $\bar{M}_{\mathrm{G}}:=\bar{M}^1_{\mathrm{G}}$. First of all, we consider the control procedure for two-cycle-cactus: choose the blue vertices as controlled vertices, and make their unique in-neighbors as green vertices respectively.
\begin{figure}[!ht]
\centering
\includegraphics[width=0.35\textwidth=0.5]{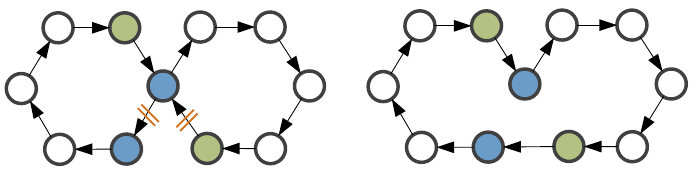}
\caption{The control of two-cycle-cactus.}\label{fig-cactus-two-leaves}
\end{figure}

For the general situation as Fig. \ref{fig-cactus-five-leaves}, we can control seven blue vertices to obtain a bigger Type II OP-CC as in the right subgraph of Fig. \ref{fig-cactus-five-leaves}. For this Type II OP-CC, since there is a directly observable vertex, it also satisfies Property $P_2$. Therefore, the right subgraph of Fig. \ref{fig-cactus-five-leaves} is an SOG.

\subsection{Observer Design for T-LGL Survival Signal Networks}
\begin{figure}[!ht]
\centering
\includegraphics[width=0.4\textwidth=0.5]{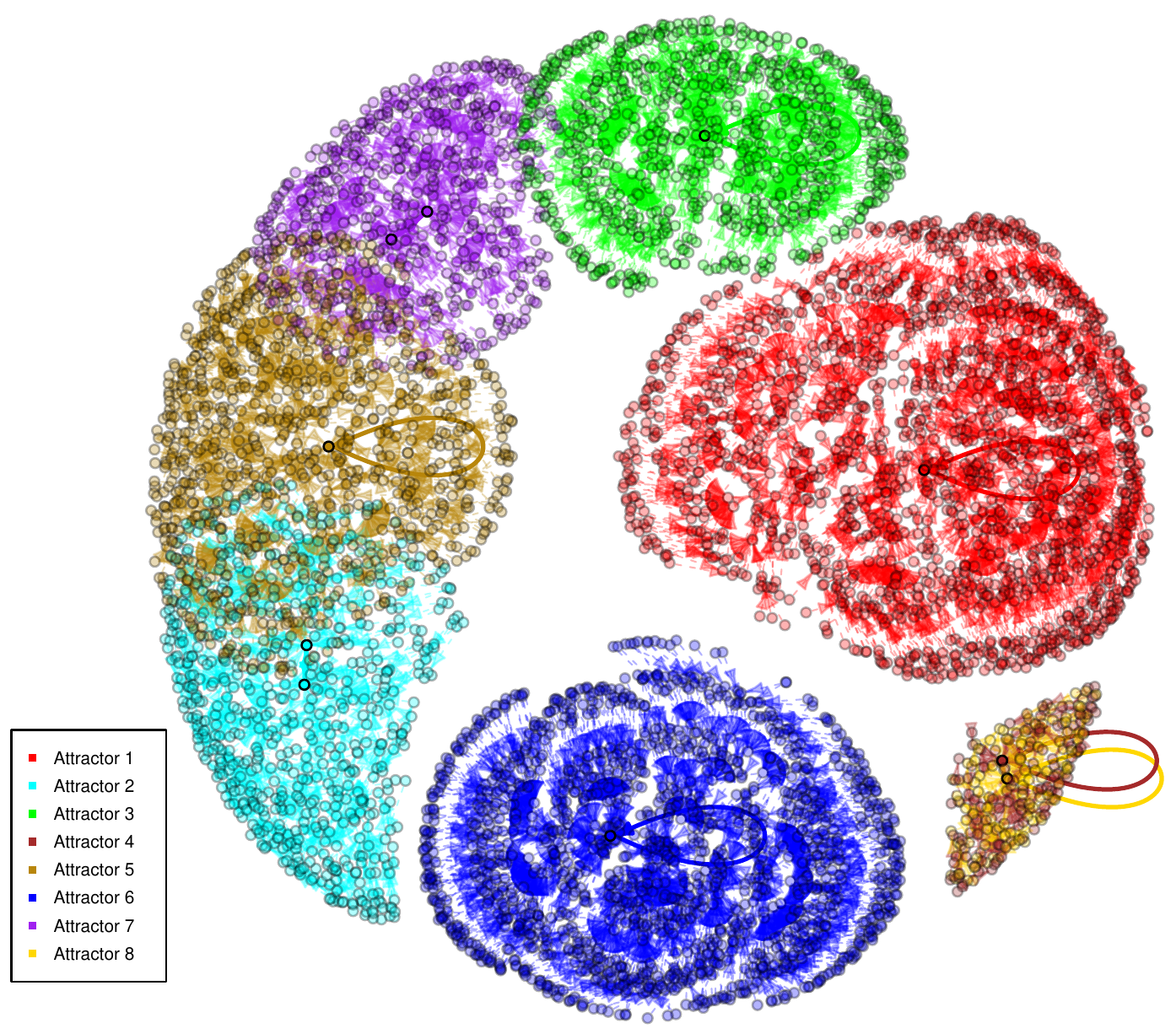}
\caption{State transition graph of Boolean network (\ref{equ-bn-29nodes}) with random $2000$ initial states.}\label{fig-stg-29nodes}
\end{figure}

This example considers the observer design problem of T-LGL survival signal network (\ref{equ-bn-29nodes}) in large granular lymphocyte leukemia (see, e.g., \cite{zhangr2008pnas16308}), where the node number is $29$. By algebraic state space representation approach, the dimension of its state transition graph will be $2^{29} \times 2^{29} \approx (5.37 \times 10^8) \times (5.37 \times 10^8)$. Although computer technology has a rapid development nowadays, this exhaustive method is still time and control dissipative. Thus, we would like to utilize the mechanism presented in subsection \ref{subsection-bn-observer} to design a feasible observer.
\begin{equation}\label{equ-bn-29nodes}
\begin{aligned}
&\text{IL15}: x_1(\#)=x_1(\ast), ~\text{RAS}: x_2(\#)=x_1(\ast),\\
&\text{ERK}: x_3(\#)=x_2(\ast), ~\text{JAK}: x_4(\#)=x_1(\ast),\\
&\text{IL2RBT}: x_5(\#)=x_1(\ast), ~\text{STAT3}: x_6(\#)=x_4(\ast),\\
&\text{IFNGT}: x_7(\#)=x_5(\ast) \vee x_6(\ast), \\
&\text{FasL}: x_8(\#)=(x_6(\ast)\wedge(x_3(\ast) \vee x_5(\ast))\vee x_{14}(\ast),\\
&\text{PDGF}: x_9(\#)=x_9(\ast), ~\text{PDGFR}: x_{10}(\#)=x_9(\ast),\\
&\text{PI3K}: x_{11}(\#)=x_{10}(\ast), ~\text{IL2}: x_{12}(\#)= \neg (x_4(\ast) \vee x_{11}(\ast)),\\
&\text{BcIxL}: x_{13}(\#)= \neg (x_4(\ast) \vee x_{11}(\ast)),\\
&\text{TPL2}: x_{14}(\#)=x_{11}(\ast),~\text{SPHK}: x_{15}(\#)=x_{11}(\ast) \vee x_{16}(\ast),\\
&\text{S1P}: x_{16}(\#)=x_{15}(\ast),~\text{sFas}: x_{17}(\#)=x_{15}(\ast),\\
&\text{Fas}: x_{18}(\#)= \neg x_{17}(\ast)\vee(\neg x_1(\ast) \wedge \neg x_{11}(\ast)),\\
&\text{DISC}: x_{19}(\#)=x_{18}(\ast), ~\text{Caspase}:x_{20}(\#)=\neg x_{1}(\ast) \wedge x_{19}(\ast),\\
&\text{Apoptosis}: x_{21}(\#)=x_{20}(\ast), ~\text{LCK}: x_{22}(\#)=x_1(\ast),\\
&\text{MEK}: x_{23}(\#)=x_2(\ast),~\text{GZMB}: x_{24}(\#)=x_4(\ast),\\
&\text{IL2RAT}: x_{25}(\#)=x_{12}(\ast),~\text{FasT}: x_{26}(\#)=x_{14}(\ast),\\
&\text{RANTES}: x_{27}(\#)=x_{14}(\ast),\\
&\text{A20}: x_{28}(\#)=x_{14}(\ast),~\text{FLIP}: x_{29}(\#)=x_{11}(\ast).
\end{aligned}
\end{equation}
and the original output measurements are assumed to be $y_1(t)=x_3(t)$, $y_2(t)=x_5(t)$ and $y_3(t)=x_6(t)$. First of all, we would like to check the observability of Boolean network (\ref{equ-DTS}). As drawn in Fig. \ref{fig-stg-29nodes}, the state transition of this Boolean network is presented by randomly choosing $2000$ initial states. For attractors $1$ and $2$, they contain the same output measurement $y_1=y_2=y_3=1$.
$$\text{attractor} ~1: 1111\vdots1111\vdots1110\vdots0111\vdots1000\vdots0111\vdots01111,$$
$$\text{attractor} ~2: 1111\vdots1111\vdots0000\vdots0011\vdots1000\vdots0111\vdots00000.$$
Hence, this Boolean network is not observable and, of course, is not structurally observable.

\begin{figure*}
\centering
\subfigure[The interaction digraph of Boolean networks (\ref{equ-bn-29nodes})]{
\includegraphics[scale=0.7]{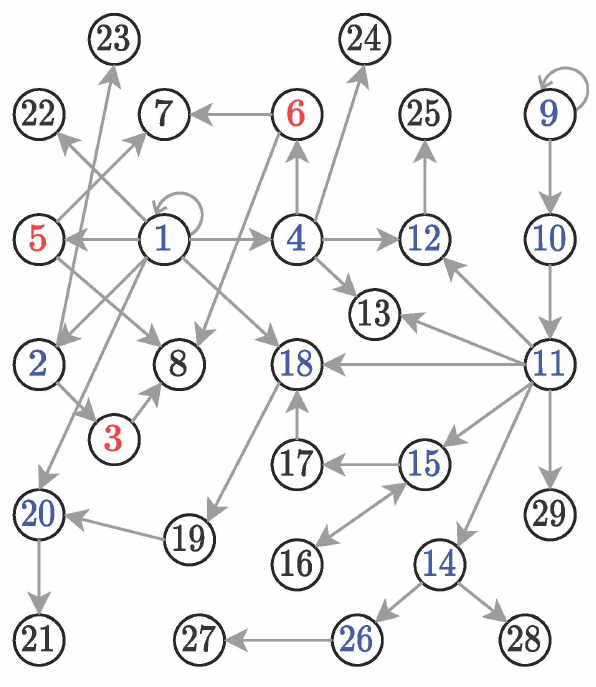}
\label{fig-29-nodes-ns}
}
\quad
\subfigure[The digraph obtained by controlling a vertex for each vertex colored by black.]{
\includegraphics[scale=0.7]{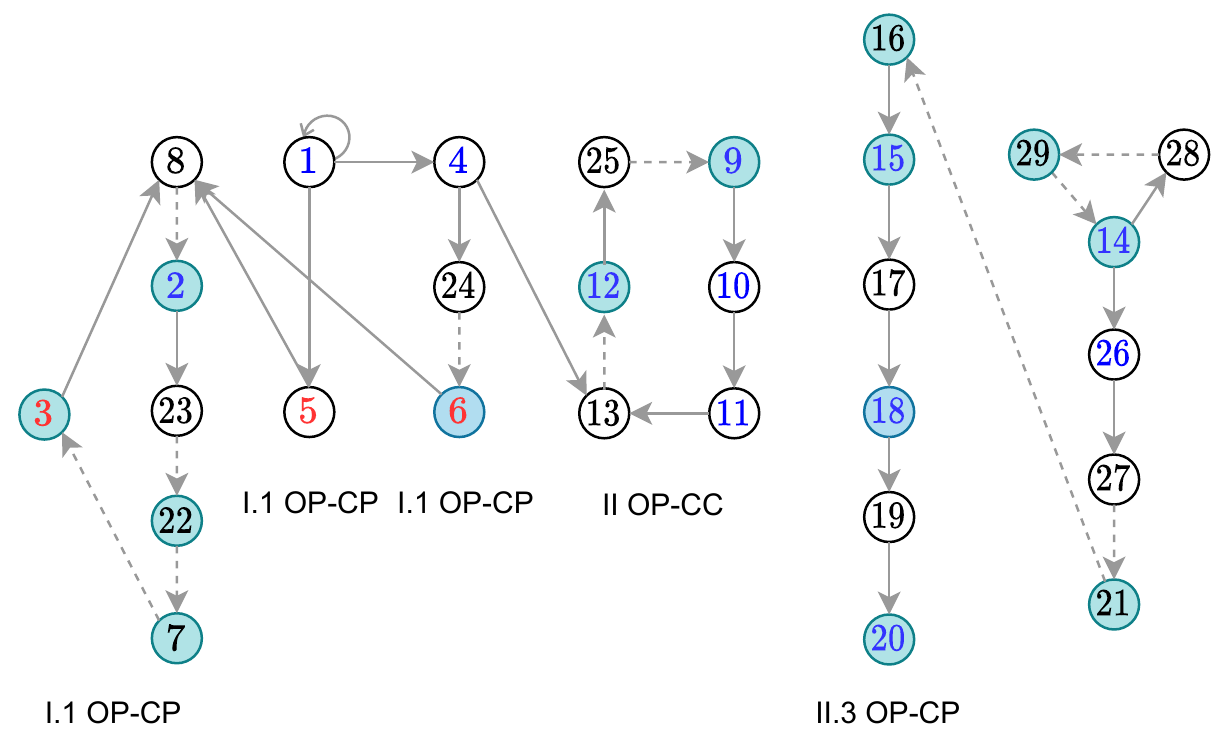}
\label{fig-29nodes-ns-control}
}
\caption{The interaction digraph of Boolean network (\ref{equ-bn-29nodes}) and the digraph after the first step.}\label{fig-exa-ns-29nodes}
\end{figure*}
Then, we select the minimum number of pinned nodes from the viewpoint of SOGs. Accordingly, its interaction digraph is drawn as the $\{3,5,6\}$-marked digraph in \ref{fig-29-nodes-ns}, where red-labelled vertices are directly observable vertices, black-labelled vertices do not satisfy Property $P_1$. That is,
$$\begin{aligned}&\bar{M}_{\mathrm{G}}:=\{v_7,v_8,v_{13},v_{16},v_{17},v_{19},v_{21},v_{22},v_{23},\\
&~~~~~~~~~~~~~~~~~~~~~~~~~~~~~~~~~~~~~v_{24},v_{25},v_{27},v_{28},v_{29}\},\\
&\bar{M}^1_{\mathrm{G}}:=\{v_{16},v_{17},v_{19}\},\\
&\bar{M}^2_{\mathrm{G}}:=\{v_7,v_8,v_{13},v_{21},v_{22},v_{23},v_{24},v_{25},v_{27},v_{28},v_{29}\}.\end{aligned}$$
Besides, one can check that there is not any OP-CC in Fig. \ref{fig-29-nodes-ns}, that is, $S_{\mathrm{G}}=\emptyset$. Hence, by Theorem \ref{theorem-all-=}, the minimum number of controlled vertices can be calculated as $\mathrm{N}^\ast_{\mathrm{G}}=\mid \bar{M}_{\mathrm{G}} \mid + \mid S_{\mathrm{G}} \mid=14$.

Subsequently, we adopt Algorithm \ref{algorithm-cyclic-P2} to design the function $\phi_v^{-}$ over $\bar{M}_{\mathrm{G}}$. As illustrated in Fig. \ref{fig-29nodes-ns-control}, we firstly select blue vertices as the controlled vertices for those in set $\bar{M}_{\mathrm{G}}$, once the desired edges are added, the original incoming edges for each controlled vertex should be deleted. Observing Fig. \ref{fig-29nodes-ns-control}, there are three Type I.1 OP-CPs $\langle v_8v_2v_{23}v_{22}v_7 v_3 \rangle$, $\langle v_1v_5 \rangle$, $\langle v_4v_{24}v_{6} \rangle$, one Type II.3 OP-CP $\langle v_{28}v_{29}v_{14}v_{26}v_{27}v_{21}v_{16}v_{15}v_{17}v_{18}v_{19}v_{20}\rangle$, as well as one Type II OP-CC $\langle v_{25}v_{9}v_{10}v_{11}v_{13}v_{12} \rangle$.

\begin{figure}[!ht]
\centering
\includegraphics[width=0.45\textwidth=0.5]{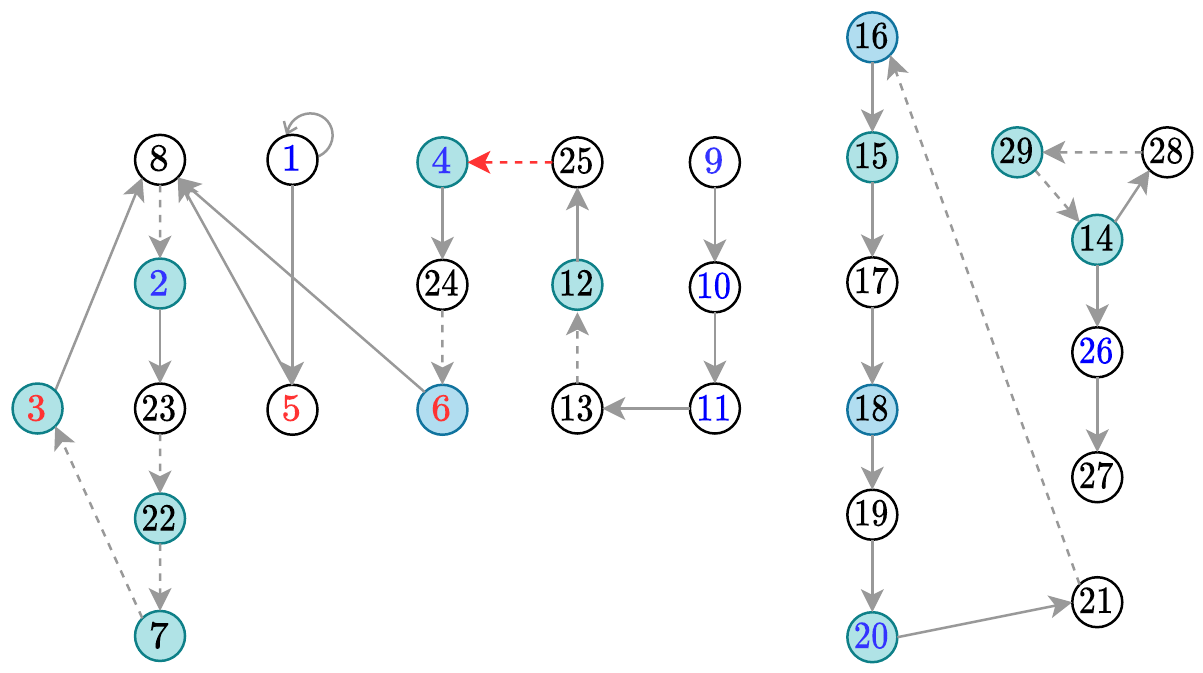}
\caption{The digraph by controlling Type II OP-CC.}\label{fig-exa-after2}
\end{figure}

\begin{figure}[!ht]
\centering
\includegraphics[width=0.45\textwidth=0.5]{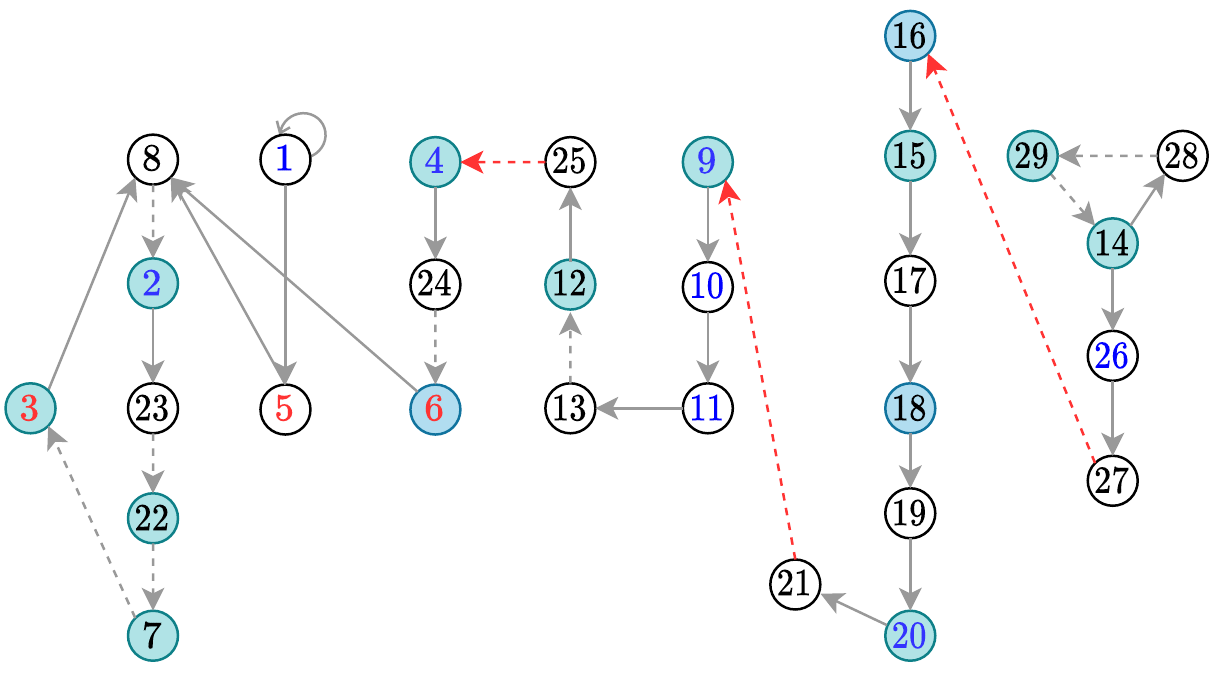}
\caption{The final returned SOG from Fig. \ref{fig-29-nodes-ns} by Algorithm \ref{algorithm-cyclic-P2}.}\label{fig-exa-sog}
\end{figure}

Finally, we consider how to combine these components to induce an SOG. For Type II OP-CC $\langle v_{25}v_{9}v_{10}v_{11}v_{13}v_{12} \rangle$, according to lines $24-26$ of Algorithm \ref{algorithm-cyclic-P2}, we replace the controlled vertex $v_9$ in this Type II OP-CC as vertex $v_4$ and add edge $(v_{25},v_4)$. To control the terminal vertex $v_{20}$, since vertex $v_{20}$ satisfies Property $P_1$ in the original interaction digraph, by lines $15-19$ of Algorithm \ref{algorithm-cyclic-P2}, we modify vertex $v_{21}$ as a non-controlled vertex (see Fig. \ref{fig-exa-after2}). For Type II OP-CC $\langle v_{16}v_{15}v_{17}v_{18}v_{19}v_{20}v_{21} \rangle$ in Fig. \ref{fig-exa-after2}, according to lines 24-26 of Algorithm \ref{algorithm-cyclic-P2}, let $\phi_{v_{21}}=v_9$. Then, via lines 20-23 of Algorithm \ref{algorithm-cyclic-P2}, let $\phi_{27}=v_{16}$. Consequently, the returned digraph (see Fig. \ref{fig-exa-sog}) is an SOG. Accordingly, the function $\phi_v$ is assigned as
$$\begin{aligned}
&\phi_{v_7}=v_3,\phi_{v_{22}}=v_7,\phi_{23}=v_{22},\phi_{v_8}=v_2,\phi_{v_{24}}=v_{6},\phi_{v_{25}}=v_4,\\
&\phi_{v_{13}}=v_{12},\phi_{v_{21}}=v_{9},\phi_{19}=v_{20},\phi_{v_{18}}=v_{19},\phi_{v_{16}}=v_{15},\\
&\phi_{v_{27}}=v_{16},\phi_{v_{29}}=v_{14},\phi_{v_{28}}=v_{29},\end{aligned}$$
and the minimal pinned node set $\Lambda^\ast$ is returned as $\Lambda^\ast=\{v_2,v_3,v_4,v_6,v_7,v_9,v_{12},v_{14},v_{15},v_{16},v_{18},v_{20},v_{22},v_{29}\}$.

By utilizing the semi-tensor product of matrices and the canonical form of logical variables, the observer can be designed as
\begin{equation}\label{equ-exa-observer}
\left\{\begin{aligned}
&\hat{x}_3[k+1]=\hat{x}_7[k],~\hat{x}_7[k+1]=\hat{x}_{22}[k],~\hat{x}_{22}[k+1]=\hat{x}_{23}[k], \\
&\hat{x}_2[k+1]=\hat{x}_8[k],~\hat{x}_6[k+1]=\hat{x}_{24}[k],~\hat{x}_4[k+1]=\hat{x}_{25}[k],\\
&\hat{x}_{12}[k+1]=\hat{x}_{13}[k],~\hat{x}_{15}[k+1]=\hat{x}_{16}[k],~\hat{x}_{16}[k+1]=\hat{x}_{27}[k],\\
&\hat{x}_{18}[k+1]=\hat{x}_{17}[k],~\hat{x}_{20}[k+1]=\hat{x}_{19}[k],~\hat{x}_{14}[k+1]=\hat{x}_{29}[k],\\
&\hat{x}_{29}[k+1]=\hat{x}_{28}[k],~\hat{x}_{9}[k+1]=\hat{x}_{21}[k],\\
&\hat{x}_i[k+1]=f_i((\hat{x}_j[k])_{j\in \mathbf{N}_i}),~v_i\not\in\Lambda^\ast,\\
&x_i[0]= \hat{f}_{\xi^{\pi_i}_{\lambda_i}} \circ \hat{f}_{\xi^{\pi_i}_{\lambda_i-1}} \circ \cdots\circ \hat{f}_{\xi^{\pi_i}_{1}}(x_{\pi_i}[\lambda_i]),\\
&\pi_i=1, ~i\in\{2,3,7,8,22,23\},\\
&\pi_i=2,~i\in\{1,5\},\\
&\pi_i=3,~i\in[1,29]_{\mathbb{N}}\backslash\{1,2,3,5,7,8,22,23\},
\end{aligned}\right.
\end{equation}
where $\hat{f}$ is the node dynamics in (\ref{equ-exa-observer}). By comparison, if we apply Algorithm 1 in \cite{zhusy2020novel} to search the pinned nodes, it is possible to let $\phi^{-}_{v_{4}}=\{v_9\}$ and $\phi^{-}_{v_{10}}=\{v_{20}\}$. In such scene, the number of pinned nodes would be $15$, which has not been minimized.

\section{Conclusion}\label{section-conclusion}
This paper addressed the structural observability analysis of discrete-time iteration by utilizing available interaction digraph while full node dynamics were unknown. Motivated by structural observability of Boolean networks, the SOGs were formalized for the first time and a polynomial-time checkable condition was also presented. Further, in order to minimize the sensor/control cost, two minimum realization problems were considered and respectively solved by two polynomial-time algorithms according to the existence of cyclic structure in the given marked digraph. Finally, some applications of SOGs were investigated to illustrate the effectiveness of our results.

As mentioned above, this work has the potential to investigate the structural observability of LSBNs, finite-field networks, and so on. In particular, the established SOGs for Boolean networks are only $n \times n$-dimensional, and the developed algorithms are all bounded by polynomial-time complexity. Thus, the obtained results will not be limited by network sizes. Moreover, the SOGs guarantee the observability of finite-field networks from the viewpoint of interaction digraphs rather than the existence of a group of system parameters. Additionally, the developed algorithms can be applied to reduce the number of pinned nodes to the lowest level. In \cite{zhusy2020novel}, the pinned nodes were searched by a polynomial-time algorithm but the node number was not minimal.

Similarly, the structurally controllable conditions, given in \cite{zhusy2020framework}, for Boolean networks can be seemed as a type of structurally controllable graphs for iteration (\ref{equ-DTS}). In this setting, the minimum realization problem of structurally controllable graphs has been proved to be {\bf NP}-hard in \cite{zhusy2020framework}. The future attention may focus on the minimum realization of SOGs with constraints. While communication are considered, it would be interesting to find out whether the minimum realization of SOGs with constraints be {\bf NP}-hard or still polynomial-time solved like conventional ones.

\section*{Appendix I: An Example to Illustrate The Structural Observability of Boolean Networks}
\begin{example}
Let each agent in (\ref{equ-DTS}) take value from the set $\mathscr{B}$, the dynamics of agents evolution is described as follows:
\begin{equation}\label{equ-exa1-BN-3nodes}
\left\{\begin{aligned}
x_1[k+1]&=x_2[k]\\
x_2[k+1]&=x_3[k]\\
x_3[k+1]&=x_1[k] \odot_1 (x_2[k] \odot_2 x_3[k])\\
y_1[k]&=x_1[k]
\end{aligned}\right.
\end{equation}
where $\odot_1,\odot_2\in\{\vee,\wedge\}$ are arbitrarily selected logical operators. Accordingly, its dependency graph $\mathrm{G}$ can be displayed as in Fig. \ref{fig-ex1-BN-3nodes}. It is obvious that there is a unique observed path $\langle v_3v_2v_1\rangle$ covering all vertices of dependency graph $\mathrm{G}$. Thus, according to Theorem \ref{theorem-stru.obser.} or Corollary \ref{corollary-stru.obser.}, this Boolean network is structurally observable.
\begin{figure}[!ht]
\centering
\includegraphics[width=0.15\textwidth=0.3]{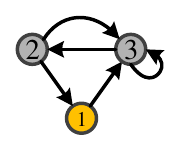}
\caption{The dependency graph $\mathrm{G}$ of Boolean network (\ref{equ-exa1-BN-3nodes}).}\label{fig-ex1-BN-3nodes}
\end{figure}

Since structural observability means that arbitrary modifications for logical couplings $\odot_1$ and $\odot_2$ would not affect system observability. To illustrate this conclusion, one can refer to the state transition graph of Boolean network (\ref{equ-exa1-BN-3nodes}) as in Fig. \ref{fig-exa1-3nodes-stg}, where $\odot_1$ and $\odot_2$ are respectively chosen as $\{\vee,\wedge\}$ in turn.
\begin{figure}[htbp]
\centering
\subfigure[State transition graph of Boolean network (\ref{equ-exa1-BN-3nodes}) with $\odot_1=\vee$ and $\odot_2=\wedge$.]{
\includegraphics[width=3.5cm]{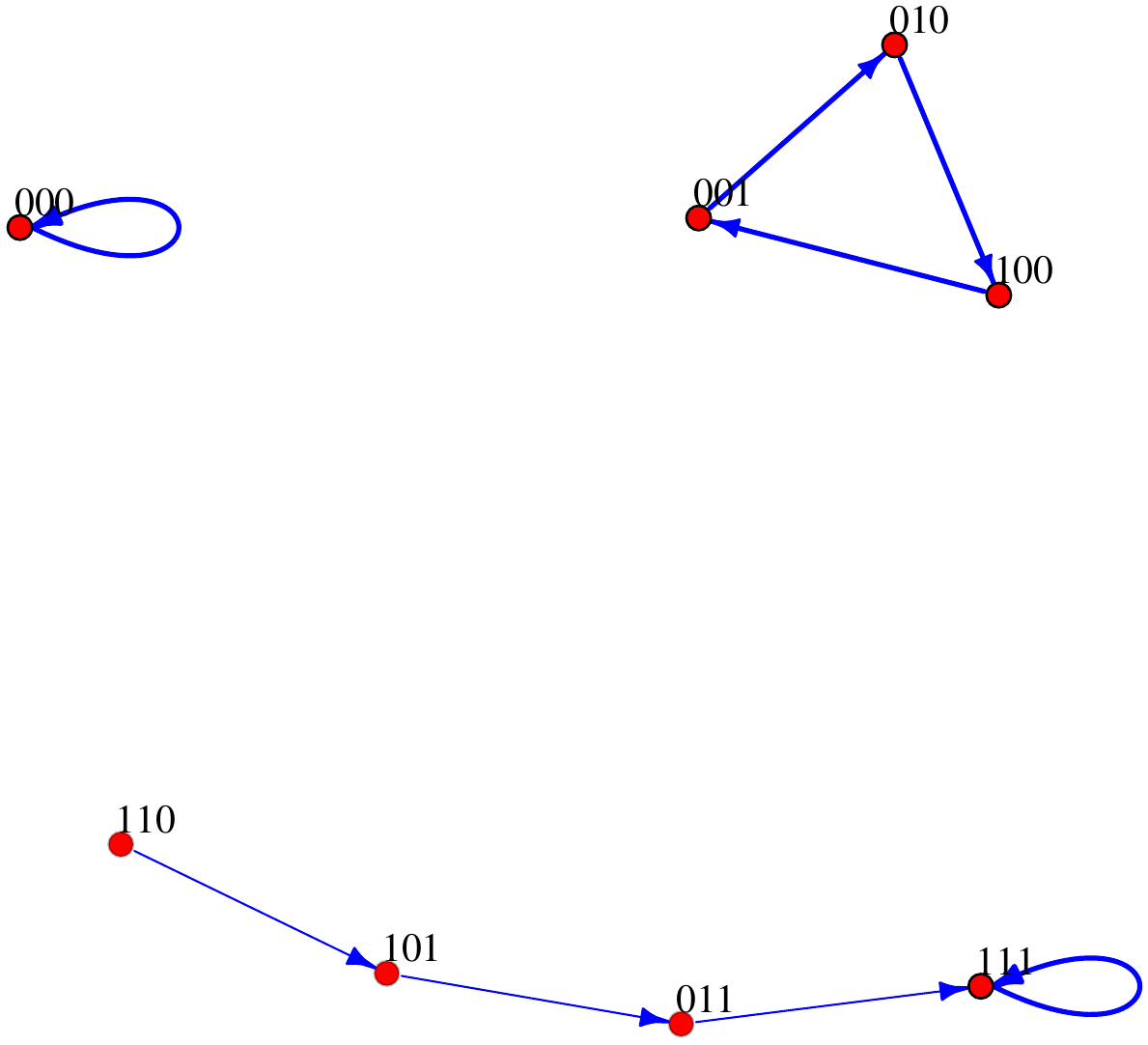}
}
\quad
\subfigure[State transition graph of Boolean network (\ref{equ-exa1-BN-3nodes}) with $\odot_1=\vee$ and $\odot_2=\vee$.]{
\includegraphics[width=3.5cm]{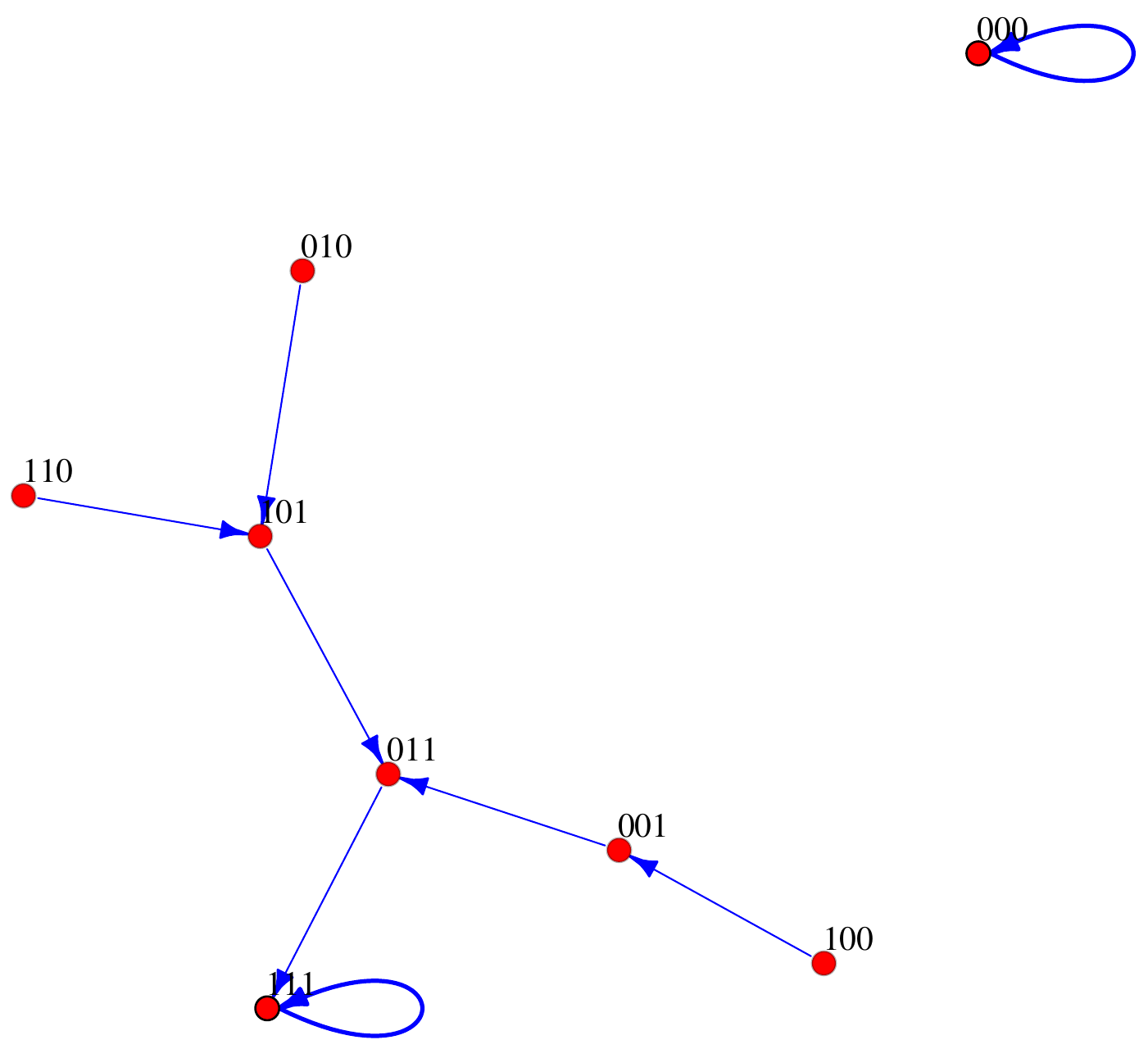}
}
\quad
\subfigure[State transition graph of Boolean network (\ref{equ-exa1-BN-3nodes}) with $\odot_1=\wedge$ and $\odot_2=\wedge$.]{
\includegraphics[width=3.5cm]{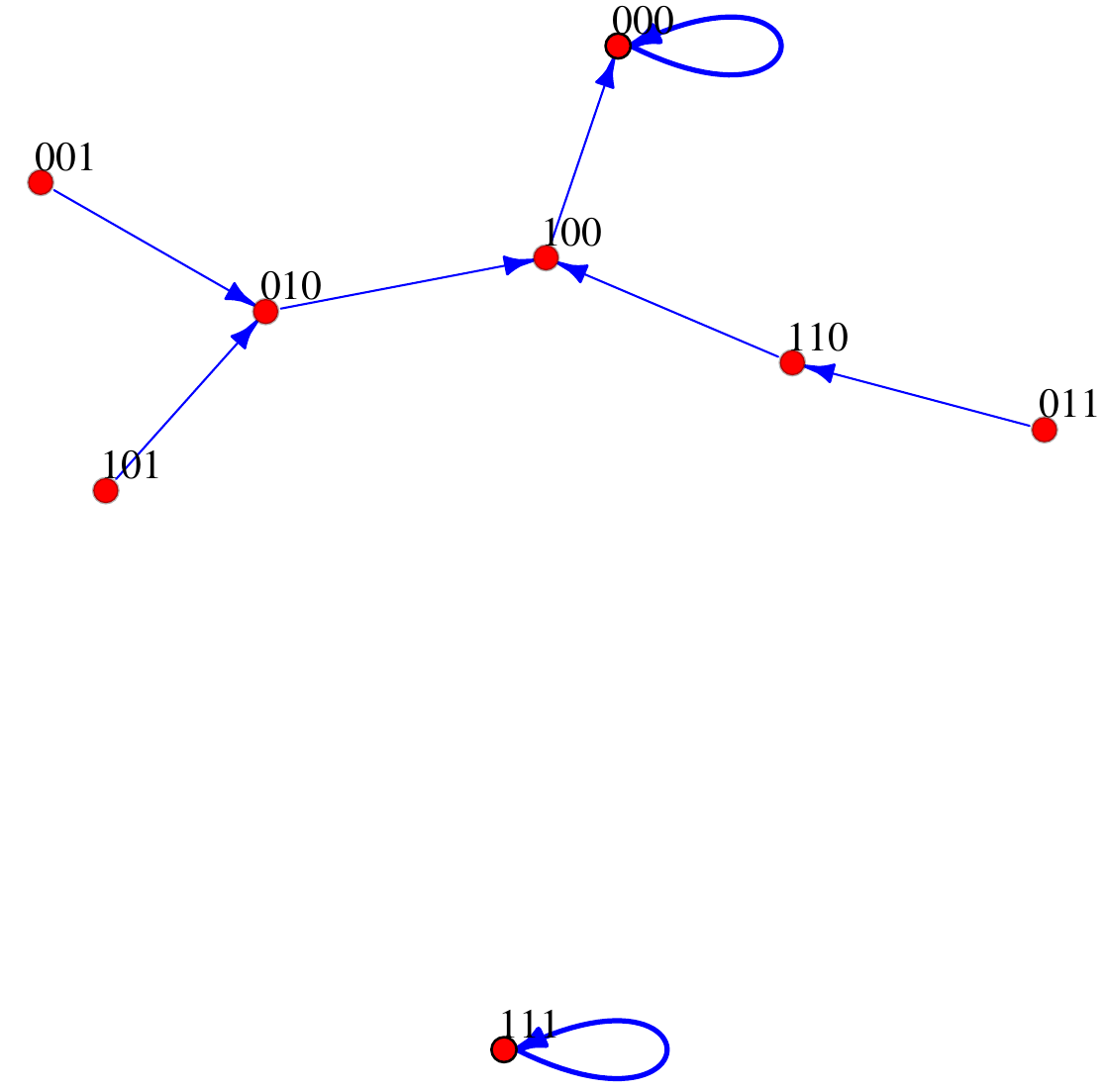}
}
\quad
\subfigure[State transition graph of Boolean network (\ref{equ-exa1-BN-3nodes}) with $\odot_1=\wedge$ and $\odot_2=\vee$.]{
\includegraphics[width=3.5cm]{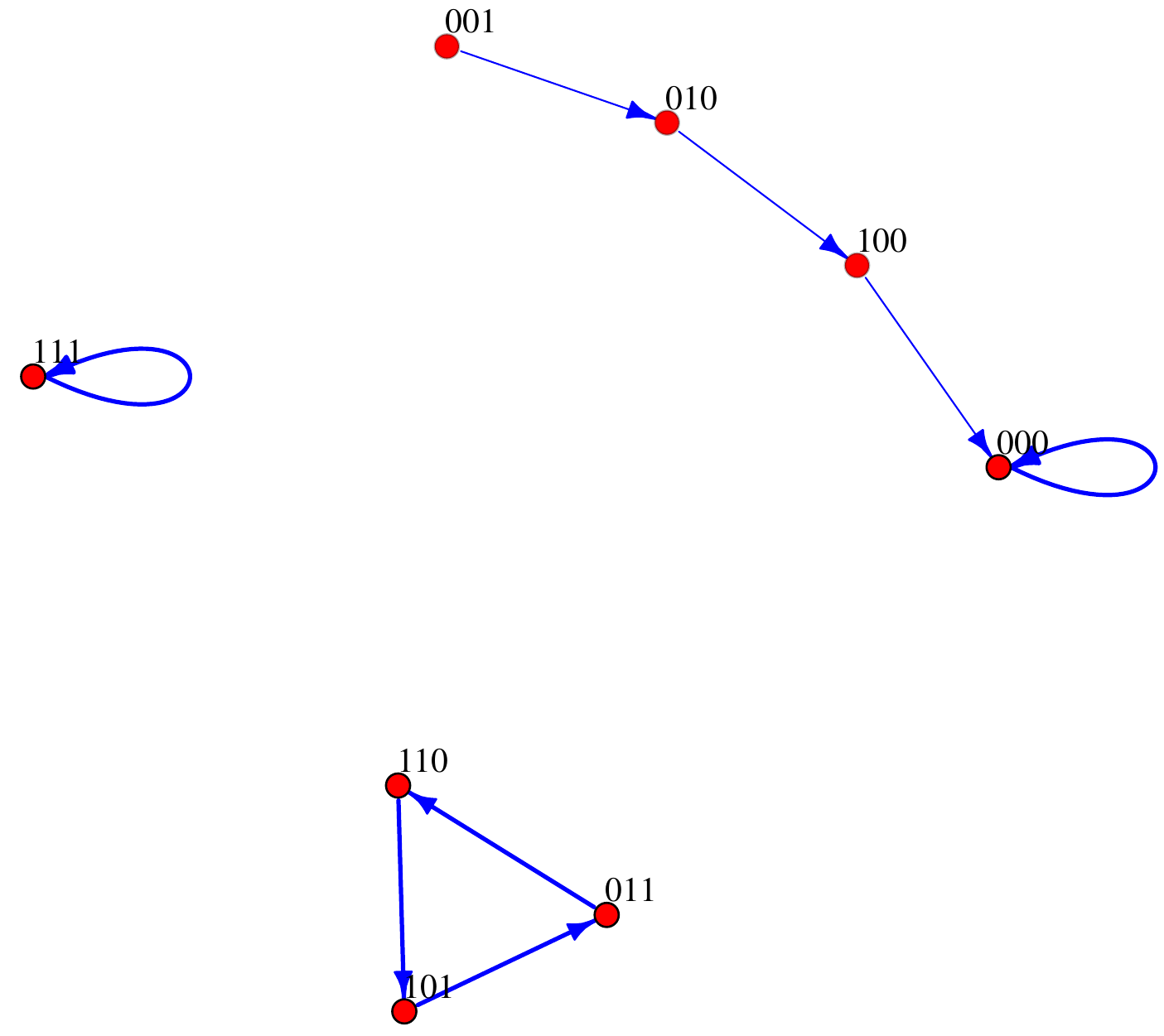}
}
\caption{State transition graph of Boolean network (\ref{equ-exa1-BN-3nodes}) with $\odot_1,\odot_2\in\{\vee,\wedge\}$.}\label{fig-exa1-3nodes-stg}
\end{figure}

It is visual that these four state transition graphs do not contain any cycle or two incoming edges of a vertex that are associated with the same output sequence. Thus, by [Theorem 6, \cite{Margaliot2013Observability2351}], these four Boolean networks are observable. This conclusion is consistent with that by Theorem \ref{theorem-stru.obser.}. The exhaustive state space approach (see, e.g., \cite{chengdz2011springer,Margaliot2013Observability2351}) is based on the state transition graph analysis in essence. Thus, one can observe from Fig. \ref{fig-exa1-3nodes-stg} that the vertices of state transition graphs for Boolean networks with $3$ nodes have reached $2^3=8$. Again, considering the logical couple number, it has generated $2^2=4$ subgraphs even if we do not consider the negation ``$\neg$''. With the increase of node number and coupling number, such brute-force analysis would gradually emerge its disadvantage, i.e., very high time complexity. By comparison, the interaction digraph utilized in Theorem \ref{theorem-stru.obser.} or Corollary \ref{corollary-stru.obser.} is only $n \times n$-dimensional, and the conditions in Theorem \ref{theorem-stru.obser.} or Corollary \ref{corollary-stru.obser.} can be checked with time $O(n^2)$. Hence, time complexity will be dramatically reduced particularly for LSBNs.
\end{example}

\section*{Appendix II: The Detailed Proof of Theorem \ref{theorem-=}}
\begin{proof}[The proof of Theorem \ref{theorem-=}]
According to Corollary \ref{corollary-geq}, one has known that $\mathrm{N}^\ast_{\mathrm{G}}\geq\mid \bar{M}_{\mathrm{G}} \mid$ holds for an arbitrarily given acyclic $[1,h]_{\mathbb{N}}$-marked digraph $\mathrm{G}$. Therefore, if we can determine $\mid \bar{M}_{\mathrm{G}} \mid$ key vertices so that controlling them can make digraph $\mathrm{G}$ become an SOG, this theorem can be naturally concluded. To this end, we will provide a detailed procedure to find these $\mid \bar{M}_{\mathrm{G}} \mid$ key vertices and develop the scheme about how to control these vertices.

With regard to vertex set $\bar{M}_{\mathrm{G}}$, we can search a set $\nabla_1(\bar{M}_{\mathrm{G}})$ of all the vertices $v_i$ satisfying that $d_{\bar{M}_{\mathrm{G}} \rhd i}=1$; the set $\nabla_1(\bar{M}_{\mathrm{G}})$ can be mathematically computed as
$$\nabla_1(\bar{M}_{\mathrm{G}})=\left\{v_j\in\mathrm{V}\mid\sum_{v_i\in \bar{M}_{\mathrm{G}}}[A_{\mathrm{G}}]_{ij} >0\right\}.$$
Correspondingly, we then construct the following undirected graph $\tilde{\mathrm{G}}:=(\tilde{\mathrm{V}},\tilde{\mathrm{E}})$. The vertex set $\tilde{\mathrm{V}}$ consists of the vertex set $\nabla_1(\bar{M}_{\mathrm{G}})\cup \bar{M}_{\mathrm{G}}$ and the duplicate $\mathrm{V}'$ of the set $\nabla_1(\bar{M}_{\mathrm{G}})\cap \bar{M}_{\mathrm{G}}$, that is, $\tilde{\mathrm{V}}:=\nabla_1(\bar{M}_{\mathrm{G}})\cup \bar{M}_{\mathrm{G}} \cup \mathrm{V}'$. Then, the edge set $\tilde{\mathrm{E}}$ is designed as follows: let $(v_i,v_j) \in \tilde{\mathrm{E}}$ if $(v_i,v_j)\in\mathrm{E}$ with $v_i \in \bar{M}_{\mathrm{G}}$ and $v_j \in \nabla_1(\bar{M}_{\mathrm{G}})\backslash\bar{M}_{\mathrm{G}}$, and $(v_i,v'_j) \in \tilde{\mathrm{E}}$ if $(v_i,v_j)\in \mathrm{E}$ with $v_i \in \bar{M}_{\mathrm{G}}$ and $v_j \in \nabla_1(\bar{M}_{\mathrm{G}})\cap\bar{M}_{\mathrm{G}}$, where we also require that $v_i\neq v_j$. Accordingly, the constructed graph $\tilde{\mathrm{G}}$ is a bipartite graph. As we can color the vertices in $\bar{M}_{\mathrm{G}}$ and $(\nabla_1(\bar{M}_{\mathrm{G}})\backslash\bar{M}_{\mathrm{G}})\cup \mathrm{V}'$ respectively by blue and red, as a consequence there does not exist an edge $(v_i,v_j)\in\tilde{\mathrm{E}}$, which connects with the same color vertices $v_i$ and $v_j$.
\begin{figure}[!ht]
\centering
\includegraphics[width=0.48\textwidth=0.5]{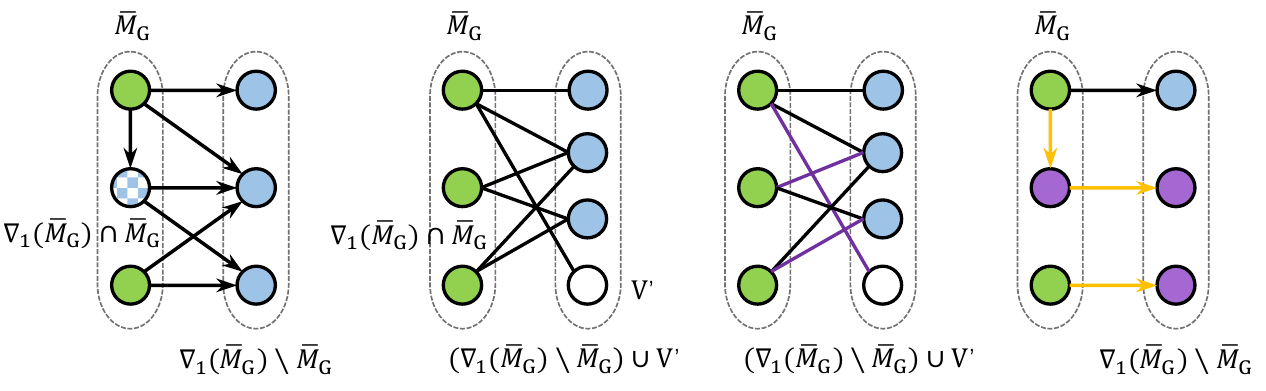}
\caption{The above graph visualizes the procedure to control the essential vertices. The first subgraph is induced by the vertex set $\bar{M}_{\mathrm{G}} \cup \triangledown_1(\bar{M}_{\mathrm{G}})$, where the vertices in left column and right columns respectively the vertices in $\bar{M}_{\mathrm{G}}$ and $\nabla_1(\bar{M}_{\mathrm{G}})\backslash\bar{M}_{\mathrm{G}}$, and the spotted vertices stand for the vertices in $\nabla_1(\bar{M}_{\mathrm{G}})\cap\bar{M}_{\mathrm{G}}$. The second subgraph shows the construction of undirected bipartite graph $\tilde{\mathrm{G}}:=(\tilde{\mathrm{V}},\tilde{\mathrm{E}})$. Thereinto, the duplicate vertex set $\mathrm{V}'$ of $\triangledown_1(\bar{M}_{\mathrm{G}}) \cap \bar{M}_{\mathrm{G}}$ is added as white vertices, and the arc $(v_i,v_j)$ with $v_i\in \bar{M}_{\mathrm{G}}$ and $v_j\in \triangledown_1(\bar{M}_{\mathrm{G}}) \cap \bar{M}_{\mathrm{G}}$ corresponds to the undirected edge connecting $v_i$ and $v'_j$, where the blue vertices belong to vertex set $\triangledown_1(\bar{M}_{\mathrm{G}}) \backslash \bar{M}_{\mathrm{G}}$. The third subgraph provides the maximum matching of the constructed graph $\tilde{\mathrm{G}}$, where the purple edges are the matching edges. The purple vertices in the forth subgraph stand for the controlled vertices and the yellow edges are the desired in-neighbor adjacency relation for each controlled vertex.}\label{fig:bipartite.graph}
\end{figure}

As for the above bipartite graph $\tilde{\mathrm{G}}$, we can apply the Hopcroft-Krap algorithm to find its maximum matching $\chi^{\ast}_M(\tilde{\mathrm{E}})$. Then, we use edge set $\chi^{\ast}_M(\tilde{\mathrm{E}})$ to induce the subgraph $\chi^{\ast}_M(\tilde{\mathrm{G}}):=(\chi^{\ast}_M(\tilde{\mathrm{V}}),\chi^{\ast}_M(\tilde{\mathrm{E}}))$ of digraph $\tilde{\mathrm{G}}$. Further, this theorem is respectively discussed in the following three situations.

(1) The first case is $\frac{\mid\chi^\ast_M(\tilde{\mathrm{V}})\mid}{2} = \mid\bar{M}_{\mathrm{G}}^2\mid$ and $\bar{M}^1_{\mathrm{G}}=\emptyset$. If such scene occurs, for every $v_k\in \bar{M}^2_{\mathrm{G}}$, we can find the unique vertex $v_j\in \chi^\ast_M(\tilde{\mathrm{V}})$ such that $(v_j,v_k)\in \chi^\ast_M(\tilde{\mathrm{E}})$. Then, one can pick the vertex set $((\nabla_1(\bar{M}_{\mathrm{G}}) \backslash \bar{M}_{\mathrm{G}})\cap \chi_M^\ast(\tilde{\mathrm{V}})) \cup \{ v_j \mid v_j'\in \chi^\ast_M(\tilde{\mathrm{V}}) \cap \mathrm{V}'\}$ as the minimum vertex set $\Lambda^\ast$ in Problem \ref{problem-P2}, and modify their in-neighbors as follows:
\begin{itemize}
  \item[$\ast$] For each vertex $v_j\in (\nabla_1(\bar{M}_{\mathrm{G}})\backslash\bar{M}_{\mathrm{G}}) \cap \chi^\ast_M(\tilde{\mathrm{V}})$, let $\phi^{-}_{v_j}=v_k$ that satisfies $(v_k,v_j)\in\chi^\ast_M(\tilde{\mathrm{E}})$ and remove all incoming edges of $v_j$;
  \item[$\ast$] Consider vertex $v_j\in \{v_j \mid v_j'\in \chi^\ast_M(\tilde{\mathrm{V}}) \cap \mathrm{V}'\}$. Let $\phi^{-}_{v_j}=v_k$ satisfying $(v_k,v'_j)\in\chi^\ast_M(\tilde{\mathrm{E}})$ and remove all other incoming edges of $v_j$.
\end{itemize}

Therefore, the obtained digraph $\hat{\mathrm{G}}$ has satisfied Property $P_1$, since every vertex $v_k\in \bar{M}_{\mathrm{G}}^2$ would have an out-neighbor $\phi_{v_k}$ satisfying that $(v_k,\phi_{v_k})\in \hat{\mathrm{E}}$ and vertex $v_k$ is its unique in-neighbor. Additionally, because digraph $\hat{\mathrm{G}}$ is acyclic, digraph $\hat{\mathrm{G}}$ also can be deemed to satisfy Property $P_2$, thus it claims that digraph $\hat{\mathrm{G}}$ is an SOG. Finally, we only need to verify that $\mid ((\nabla_1(\bar{M}_{\mathrm{G}})\backslash \bar{M}_{\mathrm{G}}) \cap \chi^\ast_M(\tilde{\mathrm{V}})) \cup \{ v_j \mid v_j'\in \chi^\ast_M(\tilde{\mathrm{V}}) \cap \mathrm{V}'\}\mid = \mid\bar{M}_{\mathrm{G}}^2\mid$, which is supported by
\begin{equation}\begin{aligned}
\mid ((\nabla_1&(\bar{M}_{\mathrm{G}}) \backslash \bar{M}_{\mathrm{G}}) \cap\chi^\ast_M(\tilde{\mathrm{V}})) \cup \{ v_j \mid v_j'\in \chi^\ast_M(\tilde{\mathrm{V}}) \cap \mathrm{V}'\} \mid \\
&=\mid((\nabla_1(\bar{M}_{\mathrm{G}})\backslash \bar{M}_{\mathrm{G}})\cup\{v_j \mid v_j'\in\mathrm{V}'\}) \cap \chi^\ast_M(\tilde{\mathrm{V}})\mid \\
&=\frac{\mid \chi^\ast_M(\tilde{\mathrm{V}}) \mid}{2}= \mid\bar{M}_{\mathrm{G}}^2\mid = \mid\bar{M}_{\mathrm{G}}\mid.
\end{aligned}\end{equation}
Therefore, the conclusion $\mathrm{N}^\ast_{\mathrm{G}}=\mid \bar{M}_{\mathrm{G}} \mid$ has been established for this situation and the controlled vertices are also sought.

(2) Consider the second case of $\frac{\mid\chi^\ast_M(\tilde{\mathrm{V}})\mid}{2} = \mid\bar{M}_{\mathrm{G}}^2\mid$ and $\bar{M}^1_{\mathrm{G}}\neq\emptyset$. In the same manner as Case (1), we first control the vertices in set $((\nabla_1(\bar{M}_{\mathrm{G}})\backslash\bar{M}_{\mathrm{G}}) \cap \chi^\ast_M(\tilde{\mathrm{V}})) \cup \{ v_j \mid v_j'\in \chi^\ast_M(\tilde{\mathrm{V}}) \cap \mathrm{V}'\}$ and establish the digraph $\vec{\mathrm{G}}$ whose basic components are presented as Fig. \ref{fig-acyclic-I-II}. Note that in the above step we have controlled $\mid \bar{M}_{\mathrm{G}}^2 \mid$ vertices. Without loss of generality, we assume that there is a unique Type I OP-CP in digraph $\vec{\mathrm{G}}$.

Next, we first suppose that only one Type II OP-CP exists in digraph $\vec{\mathrm{G}}$, that is, $\bar{M}^1_{\mathrm{G}}=\{v_{\mathrm{II,t}}\}$, as described in the left subgraph of Fig. \ref{fig-acyclic-I1-II1}. Then, the remain situations could be viewed as an extension. At this time, one can choose vertex $v_{\mathrm{I,g}}$ as a new controlled vertex and dominate vertex $v_{\mathrm{I,g}}$ by vertex $v_{\mathrm{I,t}}$ to obtain digraph $\hat{\mathrm{G}}$. Noting that vertex $v_{\text{I},g}$ is selected as a controlled vertex, thus the oriented arcs incoming $v_{\text{I},g}$ in digraph $\mathrm{G}$ can be ignored. Additionally, as vertex $v_{\mathrm{I,g}}$ is the starting vertex on the Type I OP-CP, it cannot have been controlled in digraph $\vec{\mathrm{G}}$. Thereby, $\mid \bar{M}_{\mathrm{G}}^1 \mid+\mid \bar{M}_{\mathrm{G}}^2 \mid$ vertices have been controlled.

Then, we prove that the constructed digraph $\hat{\mathrm{G}}$ in the right subgraph of Fig. \ref{fig-acyclic-I1-II1} is an SOG. Since vertex $v_{\text{II,t}}$ is the unique vertex in digraph $\vec{\mathrm{G}}$ that does not satisfy Property $P_1$ but uniquely dominates vertex $v_{\text{I,g}}$ in $\hat{\mathrm{G}}$, digraph $\hat{\mathrm{G}}$ will satisfy Property $P_1$ obviously. In addition, because arc $(v_{\text{II,t}},v_{\text{I,g}})$ is added in digraph $\hat{\mathrm{G}}$, there would be an oriented cycle $\langle v_{\mathrm{I},g}v_{\mathrm{II,s}}v_{\mathrm{II},t} \rangle$ as depicted in the right subgraph of Fig. \ref{fig-acyclic-I1-II1} if $(v_{\mathrm{I},g},v_{\mathrm{II},s})\in\mathrm{E}$ holds. Otherwise, digraph $\hat{\mathrm{G}}$ being acyclic implies the satisfactory of Property $P_2$. Apparently, this cycle satisfies Property $P_2$, because vertex $v_{\text{I},h}\not\in \{v_{\text{I},g},v_{\text{II},s},v_{\text{II,t}}\}$ and vertex $v_{\text{I},g}$ is the unique in-neighbor of vertex $v_{\text{I},h}$. Thus, the constructed digraph $\hat{\mathrm{G}}$ is an SOG.

Subsequently, the remain situation is considered, that is, there are $\mid\bar{M}^1_{\mathrm{G}}\mid>1$ Type II OP-CPs. One can pick the starting vertices of $\mid\bar{M}^1_{\mathrm{G}}\mid-1$ Type I OP-CPs as the controlled vertices and modify their in-neighbors as the terminal vertex of the next Type II OP-CPs in turn as in Fig. \ref{fig-acyclic-I1-II2}. In this setup, we can combine these $\mid\bar{M}^1_{\mathrm{G}}\mid>1$ Type II OP-CPs into an augmented Type II OP-CP by controlling $\mid\bar{M}^1_{\mathrm{G}}\mid-1$ vertices. Now there are only one Type I OP-CP and one Type II OP-CP. By controlling vertex $v_{\text{I,g}}$ in Fig. \ref{fig-acyclic-I1-II2} and making $\phi^{-}_{v_{\text{I},g}}=v_{\text{II},t}$, an SOG can be generated. For this case, the total number of control vertices $\mathrm{N}^\ast_{\mathrm{G}}$ can be calculated as $\mathrm{N}^\ast_{\mathrm{G}}=\mid\bar{M}^2_{\mathrm{G}}\mid+(\mid\bar{M}^1_{\mathrm{G}}\mid-1)+1=\mid \bar{M}_{\mathrm{G}} \mid$.
\begin{figure}[htbp]
\centering
\subfigure[The case of one Type I OP-CP and one Type II OP-CP.]{
\includegraphics[scale=0.9]{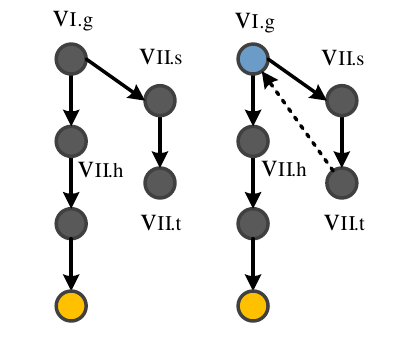}
\label{fig-acyclic-I1-II1}
}
\quad
\subfigure[The case of one Type I OP-CP and more than one Type II OP-CP.]{
\includegraphics[scale=0.9]{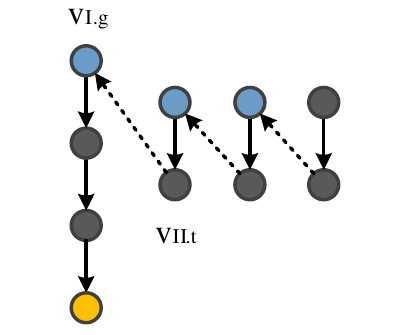}
\label{fig-acyclic-I1-II2}
}
\caption{The digraph $\vec{\mathrm{G}}$ after controlling the vertices in set $((\nabla_1(\bar{M}_{\mathrm{G}})\backslash\bar{M}_{\mathrm{G}}) \cap \chi^\ast_M(\tilde{\mathrm{V}})) \cup \{ v_j \mid v_j'\in \chi^\ast_M(\tilde{\mathrm{V}}) \cap \mathrm{V}'\}$ is presented for the second case of $\frac{\mid\chi^\ast_M(\tilde{\mathrm{V}})\mid}{2} = \mid\bar{M}_{\mathrm{G}}^2\mid$ and $\bar{M}^1_{\mathrm{G}}\neq\emptyset$.}\label{fig-acyclic-I-II}
\end{figure}

(3) With regard to the scene where $\frac{\mid\chi^\ast_M(\tilde{\mathrm{V}})\mid}{2} < \mid \bar{M}^2_{\mathrm{G}}\mid$. Denote the complementary vertex set $\bar{M}^2_{\mathrm{G}}$ by $\breve{\mathrm{V}}$. Via removing the outgoing edges from $\tilde{\mathrm{E}}$ to construct digraph $\ddot{\mathrm{G}}$, thus the maximum matching of reduced bipartite graph $\ddot{\mathrm{G}}$ satisfies $\frac{\mid \chi^\ast_M(\ddot{\mathrm{V}}) \mid}{2}=\mid \bar{M}^2_{\mathrm{G}}\backslash\breve{\mathrm{V}} \mid$. Hence, this case can be equivalently regarded as the situation with $\frac{\mid\chi^\ast_M(\ddot{\mathrm{V}})\mid}{2}$ Type I OP-CPs and $\mid \breve{\mathrm{V}} \mid +\mid \bar{M}^1_{\mathrm{G}} \mid$ Type II OP-CPs. Obviously, it is an extension of case (2). Thus, the total number of controlled vertices $\mathrm{N}^\ast_{\mathrm{G}}$ is computed as
\begin{equation}\begin{aligned}
\mathrm{N}_{\mathrm{G}}^\ast&=\frac{\mid\chi^\ast_M(\ddot{\mathrm{V}})\mid}{2} + \mid \breve{\mathrm{V}} \mid +\mid \bar{M}^1_{\mathrm{G}} \mid \\
&=\mid \bar{M}_{\mathrm{G}}^2 \backslash \breve{\mathrm{V}} \mid + \mid \breve{\mathrm{V}} \mid + \mid \bar{M}^1_{\mathrm{G}} \mid= \mid \bar{M}^2_{\mathrm{G}} \mid+\mid \bar{M}^1_{\mathrm{G}} \mid=\mid \bar{M}_{\mathrm{G}} \mid.
\end{aligned}\end{equation}

Overall, based on the analysis for above three cases, one can establish this theorem.
\end{proof}


\end{document}